\input harvmac

\def\si{\sigma}
\def\sia{{\sigma_{\!+}}}
\def\sis{{\sigma_{\!-}}}
\def\sip{{\sigma '}}
\def\sipa{{\sip_{\!\!\!+}}}
\def\sips{{\sip_{\!\!\!-}}}
\def\sipp{{\sigma'\!'}}
\def\sippa{{\sipp_{\!\!\!\!\!+}}}

\def\sippp{{\sigma'\!'\!'}}

\def\dirac{{\delta ^2}}
\def\diraccutoff{{\delta^2_\Lambda}}
\def\diraccutofftilde{{{\tilde \delta}^2_\Lambda}}

\def\minto{{\int \!\!d^2 \!\!\si \,}}
\def\mintt{{\int \!\!d^2 \!\!\si d^2 \!\!\sip \,}}
\def\DX{{\![DX]\,}} \def\Dxi{{\![D\xi]\,}}
\def\mintop{{\int \!\!d^2 \!\!\sip \,}}
\def\mintopp{{\int \!\!d^2 \!\!\sipp \,}}

\def\mbox#1#2#3{\vbox{\hrule \hbox{\vrule height#2in
          \kern#1in \vrule} \hrule\vskip#3in}}  
\def\bbox{{\mbox{0.067}{0.067}{0.01}}\;\!}
\def\const{{\rm const}} \def\nt{{\rm int}}
\def\dpmu#1{\partial_+ X^{\mu_#1}}

\def\dmmu#1{\partial_- X^{\mu_#1}}
\def\dmnu#1{\partial_- X^{\nu_#1}}
\def\dpmmu#1{\partial_+ \partial_- X^{\mu_#1}}
\def\dpm{\partial_+ \partial_-}
\def\dppmu#1{\partial^2_+ X^{\mu_#1}}
\def\dmmmu#1{\partial^2_- X^{\mu_#1}}
\def\dppmmu#1{\partial^2_+ \partial_- X^{\mu_#1}}
\def\dmmpmu#1{\partial^2_- \partial_+ X^{\mu_#1}}
\def\dpmmmu#1{\partial_+ \partial^2_- X^{\mu_#1}}
\def\dmppmu#1{\partial_- \partial^2_+ X^{\mu_#1}}
\def\m#1{{\mu_#1}} \def\n#1{{\nu_#1}}
\def\vep{{\varepsilon}}
\overfullrule=0in

\def\sitle#1#2{\nopagenumbers\abstractfont\line{#1}%
\vskip .6in\centerline{\titlefont #2}\abstractfont\vskip .3in\pageno=0}

\def\kbemail{e-mail: kbardakci@lbl.gov}
\def\jnict{Partially supported by JNICT, Lisbon.}


\def\acknowledge{This work was supported in part 
by the Director, Office of Energy
Research, Office of High Energy and Nuclear Physics, Division of High
Energy Physics of the U.S. Department of Energy under Contract
DE-AC03-76SF00098 and in part by the National Science Foundation
under grant PHY95-14797.}


\sitle{  \hfill LBNL-39854, UCB-PTH-97/03}
{\vbox{\centerline{String Field Equations}
        \vskip2pt\centerline{from}
        \vskip2pt\centerline{Generalized Sigma Model}}}
\centerline{ Korkut Bardakci\footnote{$^\sharp$}{\kbemail} and  
Luis M. Bernardo\footnote{$^\flat$}{\jnict}}
\vskip0.4cm
\centerline{\it Department of Physics\footnote{$^\dagger$}{\acknowledge}}
\centerline{\it University of California at Berkeley }
\centerline{\it and}
\centerline{\it Theoretical Physics Group}
\centerline{\it Lawrence Berkeley National Laboratory}
\centerline{\it Berkeley, CA 94720, U.S.A.}
\vskip0.4cm
\centerline{\bf Abstract }

We propose a new approach  for deriving the string field equations from a
general sigma model on the world sheet. This approach leads to an equation
which combines some of the attractive features
of both the renormalization group method and the covariant beta function
treatment of the massless excitations. It has the advantage of being
covariant under a very general set of both local and non-local transformations
in the field space. We apply it to the tachyon, massless and
first massive level, and show that the resulting field equations reproduce
the correct spectrum of a left-right symmetric closed bosonic string.

\Date{January 97}

\newsec{Introduction}

\nref\pol{J. Hughes, J. Liu and J. Polchinski, Nucl. Phys. B316 (1989) 15}
\nref\martinec{T. Banks and E. Martinec, Nucl. Phys. B294 (1987) 733}
\nref\thorn{C.B. Thorn, Phys. Rep., vol.175 (1989) 1}
\nref\bochi{M. Bochicchio, Phys. Lett. B193 (1987) 31}
\nref\witt{E. Witten, Phys. Rev. D46 (1992) 5467}
\nref\zwie{B. Zwiebach, Nucl. Phys. B480 (1996) 541 \semi
    A. Sen and B. Zwiebach, Nucl. Phys. B423 (1994) 580}
\nref\shata{S.L. Shatashvili, Phys. Lett. B311 (1993) 83}

\nref\callan{C.G. Callan, D. Friedan, E. Martinec and 
    M. Perry, Nucl. Phys. B272 (1985) 593}

\nref\buch{I.L. Buchbinder, Nucl. Phys. Proc. Suppl. B49 (1996) 133}
\nref\fuchs{U. Ellwanger and J. Fuchs, Nucl. Phys. B312 (1989) 95}
\nref\redlich{A.N. Redlich, Phys. Lett. B213 (1988) 285}
\nref\brust{R. Brustein and S. de Alwis, Nucl. Phys. B352 (1991) 45}
\nref\brustein{R. Brustein and K. Roland, Nucl. Phys. B372 (1992) 201}
\nref\tsey{A.A. Tseytlin, Int. J. Mod. Phys. A16 (1989) 4249}
\nref\vilk{I.A. Batalin and G.A. Vilkovisky, Phys. Rev. D28 (1983) 2567}

A satisfactory formulation of string field theory continues to be one of the
important open problems of string theory. There have been two main lines of
approach to this problem in the past which have enjoyed varying degrees of
success. The first approach (see refs. \refs{\thorn{--}\shata})
starts with  the BRST formalism, developed
first in the context of free strings, and generalizes it to interacting strings.
This approach was first successfully applied to the open string theory, and, 
making use of the extension of the BRST method due to Batalin and
Vilkovisky \vilk, 
it was later generalized to include closed strings \zwie. The great
advantage of this approach is that, compared to the alternatives, it is
the most developed one from the technical stand point and its correctness is
beyond doubt. However, so far it has not led to any substantial advances in
our understanding of string theory. This is no doubt due in part to the
complexity of this method, but also, it is due to the fact that initially a
fixed background has to be specified. Although there are proofs of background
independence \zwie, to our knowledge, there is no 
manifestly background independent
formulation. A great advantage of such a formulation would be its role
in unmasking various possible hidden symmetries of string theory. From the very
 beginning of string theory, symmetries like invariance under the local
transformations of the target space coordinates,
 connected with the existence of the graviton, were difficult to understand in
the usual formulation that uses a flat background metric. A
manifestly background invariant formulation of string theory should improve
our understanding of these symmetries, and also possibly shed light
 on the recently discovered symmetries such as duality.

The second main line of approach (see refs. \callan, \refs{\pol,\martinec} 
and \refs{\buch{--}\tsey}) 
to understanding string dynamics starts with a
two dimensional sigma model defined on the world sheet. 
In the earlier versions, the field content was 
restricted to massless excitations of the closed bosonic
string, namely the graviton, the antisymmetric tensor and the dilaton \callan.
The massive modes were neglected
in order to have a renormalizable and classically
scale invariant theory. The field equations are then derived by imposing
quantum scale invariance, which amounts to demanding
that the beta function vanish. In practice, one usually works with
the beta function  computed in the one loop
approximation,   using
the background field
method which preserves manifest covariance under local field
redefinitions. This approach has many advantages over the first one: The
background independence and the symmetries 
are made manifest, and the connection
between conformal invariance and renormalization is made clear. There are also
serious drawbacks: The massive modes are neglected  and an off-shell
formulation that goes beyond the equations of motion is missing. There is also
the question about what happens beyond one loop; in many important cases the
higher loop contributions lead to field redefinitions without changing the
content of the equations of motion, although it is not clear how general this
result is \tsey.

An important variation of this approach, which seems to overcome many of
the drawbacks mentioned above, is originally due to 
Banks and Martinec \martinec. The
basic idea is to apply the renormalization group equations of Wilson 
and Polchinski to the
two dimensional sigma model. The starting point is the most general sigma
model, which includes all the massive levels of the string and is
non-renormalizable in the conventional sense. A cutoff is introduced to make
the model well-defined, and the equations of motion satisfied by the string
fields are obtained by requiring the  resulting partition function
to be scale invariant. Although the  classical action is
 scale non-invariant, and the
cutoff introduces further scale breaking,  the cancellation between these
two effects  makes the final scale invariance possible. 
Hughes, Liu and Polchinski \pol\ refined 
and extended this method, and they showed that the  closed bosonic
string scattering amplitudes in the classical (tree) limit can be derived
from these equations. This approach has many nice features: It treats the
whole string all at once and not just the massless levels,
 and it is also apparently
exact and not limited to the one loop approximation. Finally,
 the emergence of the
string amplitudes as a solution provides a stringent check on the resulting
equations. Nevertheless, this approach also has some unsatisfactory features.
As already noticed in ref. \pol, the equations do not seem powerful enough to
eliminate all the unwanted states of the string spectrum; some additional
gauge invariance needed to eliminate them is apparently missing. Another
drawback is that the coordinate system in the field space is fixed right
from the beginning, and as a result,  covariance under field transformations,
which was such an attractive feature of ref. \callan, 
is lost. We suspect that these
problems are connected, and we offer some evidence in support of it.

In this paper, we propose a new approach which combines some of the
advantageous features of both the renormalization group method and the
covariant beta function treatment of the massless excitations. We start with
a general sigma model that is supposed to represent all the levels of the
closed bosonic string, with flat metric on the worldsheet.
 In section 2, the functional integral is written in
the presence of  a general background in a form completely covariant under
coordinate (field) transformations, subject only to the condition that
the determinant of the transformation is unity.
 These include not just the local
transformations associated with gravity, but also non-local transformations
which mix up different levels of the string. This covariant formulation
requires the introduction of an as yet unknown connection with the correct
transformation properties. To regulate the functional integral,
 just as in ref. \pol, we introduce
a cutoff in the free propagator, although our cutoff differs from theirs in some
details. The effective action is then required to be invariant under the
conformal (Virasoro) group in order to obtain 
the field equations. At this point,
one has to specify the variations of the fields and the cutoff under conformal
transformations. We chose the cutoff to transform exactly as in the
renormalization group method. On the other hand, the standard
expression for the conformal  generators, is not
covariant under coordinate transformations, and so
it had to be promoted to a Killing
vector with correct transformation properties. Putting everything
together, we finally arrive at our version of the 
renormalization group equations written in covariant form (see (2.37)).
This equation is not very useful as it stands; for one thing, it is an equation
in two variables, and also it has an unknown function in it. From this equation,
however,  one can derive an infinite set of equations in a single variable
 free of the
unknown function. The first of these is a one loop result, which is given
explicitly (by (2.38)), and which forms the basis of all the following work.
The remaining equations correspond to higher loops and they will not be
considered in this paper. We stress that, in our approach, one loop result is
exact; higher loops can provide more information, but they do not 
modify the one loop result \tsey.

Eq. (2.38) is still too formal to be useful as it stands; one needs explicit
results for the connection and the Killing vector. At the moment, we do not
have an exact expression for either of these, and to make progress, we resort
to an expansion which we call the quasi-local expansion. This is an expansion
in the number of derivatives on the world sheet and it is explained in section
3. The local coordinate transformations associated with gravity appear at
zeroth order, and each new power of the expansion parameter $b$ brings in
 transformations with two more derivatives on the world sheet. The levels of
the string can also be similarly organized; the nth level goes with the power
$b^{n-1}$. In section 3, we study the zeroth order term in the expansion. This
corresponds to considering only the tachyon and the massless levels and imposing
only local coordinate invariance. We are, therefore, back in familiar territory
of renormalizable sigma model \callan, where the connection  and the
generators of conformal algebra are well-known.
  Our reason for reexploring it is
twofold: Firstly, we would like to check our formalism against standard results.
This check is non-trivial since the way we treat the dilaton is different from
the standard treatment \callan, where the dilaton field is introduced as an
independent field in the action from the beginning. In our approach, the
determinant of the metric plays the role of the dilaton field, and everything
works out alright. We note that this determinant cannot be gauged away, since
the coordinate transformations under which the model is invariant
 are restricted to have unit determinant.
 The second question we would like to answer is what happens
if we abandon covariance by, for example, setting the connection equal to
zero. In this case, we recover the gravitational equations in a fixed gauge, but
we loose the equation of motion for the dilaton. Therefore, covariance is
important in obtaining a complete set of equations.

The next step is to go to first order in the expansion, which is the subject of
section 4. The first massive level of the string enters at this order, and also
the coordinate (field) transformations include non-local terms for the first
 time. The important question is whether in this case,
 a suitable metric and a Killing vector that generates the conformal algebra
exist. We show how to construct both the metric and the Killing vector to this
order, and we derive the resulting equations of motion for the first massive
level. An important check on the method is to find out whether the level
structure agrees with that of the first massive level of the string. Again, to
see what difference covariance makes, we check this for the non-covariant
version, when the connection is set equal to zero. Just as in the case of
ref. \pol, we find that there are too many states, and there is not enough gauge
invariance to eliminate the spurious states. In section 5, we investigate the
first massive level in the covariant case. Here, the situation is the opposite;
for a general left-right non-symmetric model, there are too few
states. Only when the model is left-right symmetric, there is an exact match.
We have to conclude that our approach works only for left-right symmetric
models, although at this time, we do not have a good understanding of this
restriction.

In our opinion, the main contribution of this paper is that, at least in the
context of a natural expansion, the field equations of motion that follow from
the general sigma model can be made covariant under not only local, but also
non-local transformations in the field space. Furthermore, this covariance is
crucial in eliminating spurious states of the first massive level. Clearly, as
far as this question is concerned, we
have only scratched the surface in this paper. It would be very desirable to go
beyond the expansion we have used and to establish exact covariance under
non-local transformations. A subsidiary result of this paper is the one loop
basic equation,
 which, to some extent, bridges the gap between the standard treatment of the
renormalizable sigma model, 
and the Wilson renormalization group approach. 
This equation may be of use in other applications.

\newsec{Covariant Renormalization Group Equations}

In this section, we derive a set of renormalization group equations for a
general sigma model in a classical background. These equations are
derived by imposing conformal invariance on the sigma model in the
presence  of a background field; they are covariant
generalizations of the string equations of motion derived in ref. \pol. 
Throughout, we also work with flat worldsheet.
We found the renormalization group approach of \pol\ advantageous
for the following reason: When 
the generalized sigma model action $S$  contains
all the levels of the string and not just the massless ones, one is
dealing with a conventionally non-renormalizable theory. In their
approach, the conventionally non-renormalizable interactions
coming from massive states, as well as the superrenormalizable interaction
resulting from the tachyon, are treated on equal footing with the
renormalizable interactions of the massless states.
However, there are some problems with this approach. One of them is
lack of covariance under the transformation of the target space
coordinates. For example, the equations derived
in \pol\ had a flat background; as a result, they were not explicitly
covariant even
under the usual coordinate transformations (local coordinate invariance)
associated with gravity. Also, as pointed out by them, the equations do
not seem strong enough to eliminate the states that are absent from the
string spectrum. We will overcome both of these problems, at least for the
first massive level, by combining
the renormalization group approach with
the traditional background field approach (see, for example \ref\braat
{E. Braaten, T.L. Curtright and C.K. Zachos, Nucl. Phys. B260 (1985) 630}). 
Our approach will
ensure covariance under not only local but also
arbitrary non-local coordinate transformations, and by both considering a
non-renormalizable action and also non-local coordinate transformations which
mix up levels with different masses, the traditional treatment \callan\
will be extended to include massive levels of the string.

Our starting point is the partition function
\eqn\partfunc{Z[X_{o},\Lambda]=\int \DX e^{S'[X,\Lambda]}.}
We will specify $X_{o}$ and $S'$ in terms of $X$ and the action $S$ shortly.
The action $S$, which is a functional of the 
string coordinate\foot{$X^{\mu\si}$
is the same as $X^\mu(\si)$. In this paper $\si$ always stands for worldsheet
coordinates. All other greek indices refer to spacetime coordinates.} 
$X^{\mu\si}$
and a function of the cutoff parameter $\Lambda$, can be written as
\eqn\action{\eqalign{S[X,\Lambda]&=\minto {\cal L}(X,\Lambda) \cr
        &=   X^{\mu\si}\Delta_{\mu\si,\nu\sip}
        (\Lambda)X^{\nu\sip}+ S_{\nt}[X]. \cr}}
The cutoff appears only in the quadratic part of the action through the
regularized free  inverse propagator $\Delta$;
$S_{\nt}$ is independent of the cutoff.
When not essential, we will suppress the dependence on the cutoff; later,
the cutoff dependence will be specified more precisely.

The primary goal of this paper is to formulate the string field
equations in a form covariant under arbitrary functional transformations
of the background field $X_{o}$:
\eqn\diff{X_o^{\mu\si}\rightarrow  F^{\mu\si}(X_o).}

We shall adopt the usual language of differential geometry: Tensors will be
labeled by a composite index like $\mu\si$, and upper and lower indices
will undergo the standard transformations  of contravariant and covariant
tensor indices. Also, when no confusion can arise, we follow
the convention of summation over repeated discrete indices $\mu\nu$, and
integration over repeated continuous indices $\si,\sip$. Here, $\si$ stands
for the  worldsheet coordinates $\sigma_{0}$ and $\sigma_{1}$; the worldsheet
metric is Euclidean. In the
standard background field method, it is convenient to define a new coordinate
variable
$X^{\mu\si}(s)$ as a function of an internal 
parameter $s$ through the geodesic equation
\eqn\geodes{{d\over ds} X^{\mu\si}(s)+ 
    \Gamma^{\mu\si}_{\alpha\sip,
    \beta\sipp}(X) {d\over ds} X^{\alpha\sip}(s)
    {d\over ds} X^{\beta\sipp}(s)=0,}
with the boundary condition that, at $s=1$, $X^{\mu\si}(s=1)\equiv X^{\mu\si}$,
where $X$ is the original variable that appears in \partfunc. As in this case,
when the parameter $s$ is omitted, this will mean $X$ at $s=1$.
The classical background field $X_o$ is given by
$X^{\mu\si}(s=0)\equiv X_{o}^{\mu\si}$, and it is also useful to define the
tangent
at $s=0$ by $({d X^{\mu\si}(s)}/{d s})_{s=0}\equiv \xi^{\mu\si}$. The
connection $\Gamma$ is yet unspecified; it is introduced in order to have
covariance under \diff.
We shall see later on that quantum corrections break this group down to
transformations with unit functional determinant:
\eqn\deter{\det\left({\delta F ^{\mu\si}\over \delta X^{\nu\sip}}\right)=1.}
The idea of the background field method is to change variables in \partfunc\
from $X=X(1)$ to $\xi$ at fixed $X_o$ in order to exhibit the dependence on
the classical field explicitly. This  is conveniently done by expanding $X$ and
also the action in powers of the parameter $s$ and setting $s=1$ at the end.
For later use, here we write      
down the first three terms of the expansion of $X$:
\eqn\expandX{X^{\mu\si}=X^{\mu\si}_o + \xi^{\mu\si} -{1 \over 2}
    \Gamma^{\mu\si}_{\alpha\sip,\beta\sipp}(X_o)\xi^{\alpha\sip}
    \xi^{\beta\sipp}
    + \cdots}
In the same way, the action can be expanded:
\eqn\expandS{\eqalign{S[X]&=S[X_o]+
    \left. {d \over ds} S[X(s)]\right|_{s=0}+
    {1 \over 2}\left. {d^2 \over ds^2} S[X(s)]\right|_{s=0}+ \cdots \cr
    &=S[X_o]+  {{\delta S[X_o]} \over {\delta
    X_o^{\alpha \si}}}\xi^{\alpha \si} +{1\over2}
    G_{\alpha\si,\beta\sip}(X_{o})\xi^{\alpha\si}\xi^{\beta\sip}
    + S_{R}, \cr
    &\equiv S[X_o,\xi]. \cr}}
Here, $S_{R}$ denotes the cubic and higher order terms in $\xi$ in the
expansion  of $S$.
The propagator $G$ in the presence of the background field is given by
\eqn\G{G_{\alpha\si,\beta\sip}(X_o)\equiv {{\delta^2 S[X_o]}
        \over {\delta X_o^{\alpha \si}\delta X_o^{\beta \sip}}}
        -{{\delta S[X_o]} \over {\delta X_o^{\gamma \sipp}}}\Gamma
        ^{\gamma\sipp}_{\alpha\si,\beta\sip}(X_o).}
We are now ready to define $S'$: it is gotten from $S$ by subtracting the term
linear in $\xi$:
\eqn\sp{S'[X_{o},\xi]=S[X(1)]-{{\delta S[X_o]}\over {\delta X_o^{\mu\si}}}
    \xi^{\mu\si}.}
It is well-known that the transition from $S$ to  $S'$ in the background field
approach is equivalent to the introduction of a background field dependent
source.
Changing variables of integration from $X$ to $\xi$ in \partfunc, the partition
function can be written as
\eqn\newZ{\eqalign{Z[X_o]& =\int \DX e^{S'}\cr 
       & = \int \Dxi e^{(S'+{\cal M})},\cr}}
where we have defined
\eqn\jacob{\det\left({{\partial X} \over {\partial \xi}}\right) \equiv
            \exp\left({\cal M}\right).}
From now on, we will drop the subscript on $X_o$; $X$ will stand for the
classical background field, and in order to avoid confusion, the original field
$X$ will be denoted by $X(1)$.
${\cal M}$, the $\log$ of the jacobian, can be computed from \expandX; we 
write down the result to quadratic order in $\xi$:
\eqn\M{{\cal M}=-\left ({1 \over 6}R_{\beta\si, \gamma\sip} +{1 \over 2}
    D_{\beta\si}\Gamma^{\alpha\sipp}_{\alpha\sipp,\gamma\sip}\right) \xi^{
    \beta\si}\xi^{\gamma\sip}+{\cal M}_{R},}
where ${\cal M}_{R}$ is at least cubic in $\xi$.
The combination of the Ricci tensor and the covariant derivative that appears on
the right hand side of this equation is explicitly given by
\eqn\Ricci{\eqalign{{1\over 6}&R_{\beta\si,\gamma\sip}+{1\over 2}
            D_{\beta\si}\Gamma^{\alpha\sipp}_{\alpha\sipp,\gamma\sip}
            = \cr &={1\over6}\left({{\delta \Gamma
            ^{\alpha\sipp}_{\beta\si,\gamma\sip}}
            \over {\delta X^{\alpha\sipp}}}
           + {{\delta \Gamma^{\alpha\sipp}_{\gamma\sip,\alpha\sipp}}
            \over {\delta X^{\beta\si}}}
            +{{\delta \Gamma^{\alpha\sipp}_{\beta\si,\alpha\sipp}}
            \over {\delta X^{\gamma\sip}}}\right) 
            -{1\over6}\left(\Gamma^{\alpha\sipp}_{\nu\sippp,\beta\si}
            \Gamma^{\nu\sippp}_{\alpha\sipp,\gamma\sip}
            +2 \Gamma^{\alpha\sipp}_{\alpha\sipp,\nu\sippp}
            \Gamma^{\nu\sippp}_{\beta\si,\gamma\sip}\right). \cr} }
We note that
\item{a)} ${\cal M}$ is of order $\hbar$; it is a quantum 
correction to the classical action.
\item{b)} We dropped the term linear in $\xi$ in \M; this can be 
taken care of by redefining $S'$.
\item{c)} Referring to \M, we see that the first term on the right, the 
Ricci    tensor, is covariant; however, the second term, which is the covariant
derivative of the contracted connection, is not. 
If the connection is derived
from a metric, as will be the case here, we have
\eqn\ggamm{\Gamma^{\beta\sip}_{\alpha\si,\beta\sip}={1\over 2g}{\delta g\over
\delta X^{\alpha\si}}.}
Here, $g$ is the determinant of the metric. From this, one sees that this term 
is covariant only under coordinate transformations with unit determinant.
Therefore, although we have started with a fully covariant classical
formulation, quantum corrections break the full diffeomorphism group down to
transformations with unit determinant.
\bigskip

The next step in our program is to expand the partition function (see \newZ) in
a perturbation series. However, in contrast to the usual perturbation series,
each term in our series is invariant under the restricted (unit determinant)
transformations (\diff, \deter). In deriving the perturbation expansion, we
follow the standard functional approach discussed in the textbooks 
(see for example \ref\text{J.F. Donoghue, E. Golowich, B.R. Holstein, 
Dynamics of the Standard Model, Cambridge University Press, 1992.}). 
First, we define a free  partition function $Z_o$  in the absence of
interaction (except
with the external field), coupled to an external source $J$:
\eqn\ZsourceJ{\eqalign{
            Z_o[X,J]&=\int \Dxi \exp\left(S(X)+{1\over2}
            G_{\alpha\si,\beta\sip}(X)\xi^{\alpha\si}\xi^{\beta\sip}
            +i J_{\mu\si}\xi^{\mu\si}\right) \cr
            &=\exp\left(S(X)-{1\over2}\Tr \log G+{1 \over 2}
            J_{\mu\si}G^{\mu\si,\nu\sip}(X) J_{\nu\sip}\right). \cr }}
Here, $G$ with the upper indices is the inverse of $G$ with the lower indices.
The full partition function can now be written as
\eqn\Zfull{Z(X)=\exp\left(S_{I}(X,P)\right) Z_o(X,J)|_{J=0},}
where,
\eqn\SI{S_{I}(X,\xi)=S_{R}(X,\xi)+ {\cal M}(X,\xi),}
and $P$, which replaces $\xi$ as the argument of $S_{I}$ in \Zfull, is given by
\eqn\Pee{P^{\mu\si}\equiv -i{\delta \over \delta J_{\mu\si}}.}

In \Zfull, after the functional derivatives with respect to $J$ act on
$Z_o(X,J)$, $J$ is set equal to zero. Eq. \Zfull can be used as
the starting point of a perturbation expansion in powers of $S_I$.
It is easy to see that the invariance under the coordinate
transformations \diff, subject to the constraint \deter, are preserved in
this expansion,
if at the same time, $J$ and   $P$ are transformed by
\eqn\trans{J_{\mu\si}\rightarrow \left({\delta F^{\nu\sip} \over \delta X^{
\mu\si}}\right) J_{\nu\sip}, \qquad
P^{\mu\si}\rightarrow \left({\delta F^{\mu\si}\over \delta X^
{\nu\sip}}\right)P^{\nu\sip}.}

Since we are dealing with a non-renormalizable interaction, the series is 
badly divergent. To have a well defined answer, we introduce a cutoff in 
the quadratic term in the action (see \action). 
This cutoff in general violates 
the coordinate invariance described above. We shall later see how to deal
with this problem; in fact, the solution will be at the heart of the derivation
of the string equations.

Among the coordinate diffeomorphisms, conformal transformations on the world
sheet will play a special role. They are given by
\eqn\conf{\sia\rightarrow f_{+}(\sia,\sis),\qquad 
    \sis\rightarrow f_{-}(\sia,\sis),}
where 
\eqn\comp{\sia\equiv \sigma_{0}+i \sigma_{1},\qquad
\sis\equiv \sigma_{0}- i \sigma_{1}.}
In what follows, to save writing, we will only exhibit the formulas 
corresponding to the $f_{+}$ transformations; the $f_{-}$ expressions can be
obtained from these by an interchange of + with $-$. The string field equations
follow from demanding that the partition function \partfunc\ be invariant under
the conformal transformations. The first thing to check is the invariance of 
the   quadratic part of the action in \action; in the absence of the cutoff,
$\Delta$ is given by\foot{Whenever $\partial_\sia$ acts on a function of only
$\si$, not $\si$ and $\sip$, it will be written as just
$\partial_+$.}
\eqn\DDelta{\Delta_{\mu\si,\nu\sip}\left(\Lambda =0\right)
= - \partial_{\sia}\partial_{\sis}\delta^{2}(
\si-\sip) \eta_{\mu\nu},}
and is conformally invariant. Here,
$\eta_{\mu\nu}$ is the flat Minkowski metric. We introduce the
cutoff by defining
\eqn\cut{\Delta_{\mu\si,\nu\sip}(\Lambda)=
    \eta_{\mu\nu}\Delta_{\si,\sip}(\Lambda).}

It will turn out to be useful to also define the following related
functions:
\eqn\invc{\eqalign{\Delta_{\si,\sip}(\Lambda)= -\partial_{\sia}
    \partial_{\sis}\diraccutoff(\si,\sip),& \qquad
    \partial_{\sia}\partial_{\sis}\Delta^{\si,\sip}(\Lambda)=
    - {\tilde\delta}^2_\Lambda(\si,\sip), \cr
    \mintop \Delta_{\si,\sip}(\Lambda) \Delta^{\sip\!,\sipp\!}(\Lambda)
    = \dirac (\si -\sipp), & \qquad
    \mintop {\tilde\delta}^2_\Lambda(\si,\sip) \diraccutoff(\sip,\sipp)
    = \dirac (\si -\sipp).\cr}}
The detailed structure of these functions is not important; all one needs
to know is that $\diraccutoff(\si,\sip)$ and ${\tilde\delta}^2_\Lambda
(\si,\sip)$
are smoothed out versions of the two dimensional Dirac delta function, and 
they are chosen so that $\Delta_{\si,\sip}(\Lambda)$,  and as many
derivatives  of it as needed, are finite at $\si=\sip$, when the cutoff 
is finite. Notice that,
unlike the propagator used in \pol, our propagator need not vanish at $\si=
\sip$. As a result, in contrast to \pol, we shall encounter cutoff
dependent terms in our equations. In some cases, these can be eliminated
by renormalizing, for example, the slope parameter. In other cases, when
such a renormalization is not possible, we will consider it as an anomaly
and set its coefficient equal to zero. This will then provide
additional useful information. For example, the field
equation for the dilaton is derived in this fashion.

The cutoff violates conformal invariance;
$\Delta$ with cutoff is no longer conformal invariant. To restore the
conformal invariance, we have to supplement the transformations \conf\ by a
suitable variation of the cutoff parameter(s). Specializing to infinitesimal
variations, we define
\eqn\var{\delta = \delta_\Lambda + \delta_v,}
where     $\delta_{v}$ is a ``+'' infinitesimal  conformal transformation,
which corresponds to taking the $F$ in \diff\ and \trans\ to be\foot{Then
$\delta_v=\minto v(\sia)\partial_+ X^{\mu\si}{\delta\over \delta X^
{\mu\si}}$, but in general $\delta_v=F_v^{\mu\si}(X){\delta\over \delta X^
{\mu\si}}$. To be more precise, this $F$ is not the same defined in \diff\
and \trans, but is the $\tilde{F}$ defined by $F^{\mu\si}(X)=X^{\mu\si}+
\tilde{F}^{\mu\si}(X)$, with the tilde dropped.}
\eqn\spec{F^{\mu\si}(X)\rightarrow F^{\mu\si}_{v}(X)=
    v(\sia) \partial_{+}X^{\mu\si},}
with a similar expression for the ``$-$'' transformations. Here $v$ is an 
arbitrary function of $\sia$, parametrizing conformal transformations. The
variation $\delta_{\Lambda}$ is defined so that the quadratic part of the 
action in \action\ is invariant under the total variation $\delta$, resulting
in the equation
\eqn\delflat{\partial_{\sia} \left(v(\sia)\Delta_{\si,\sip}(\Lambda)
    \right)+\partial_{\sipa}\left(v(\sipa)\Delta_{\si,\sip}(\Lambda)\right)
    -\delta_{\Lambda}\left(\Delta_{\si,\sip}(\Lambda)\right)=0.}

Later on, we will also need the cutoff variation  of the propagator, so the
variation of the function $\Delta^{\si,\sip}\!(\Lambda)$, which is the inverse
of $\Delta_{\si,\sip}(\Lambda)$, is needed. At first, it may seem that the 
inverse also satisfies the same equation; and this would be true if the 
inverse were unique. However, there is a well-known ambiguity in going from 
$\Delta$ to  its inverse; for example, in the absence of cutoff
\eqn\prop{\Delta^{\si,\sip}\!\left(\Lambda=0\right)= {1 \over {4\pi}}\log\left(
    (\si-\sip)^{2}\right)+ k_{+}(\sia) +k_{-}(\sis).}
The functions $k_{+}$ and $k_{-}$ are arbitrary, resulting in a non-unique
inverse.  The variation of the propagator under the change of
the cutoff also suffers from the same ambiguity. This ambiguity can be resolved
by demanding that the ultraviolet cutoff does not change the long distance
behavior of the propagator\foot{For a detailed treatment of this question, see
\pol.}. Imposing this boundary condition, we have the following equation:
\eqn\invflat{v(\sia)\partial_{\sia}\Delta^{\si,\sip}\!(\Lambda) +
    v(\sipa)\partial_{\sipa}\Delta^{\si,\sip}\!(\Lambda)- 
    \delta_\Lambda\Delta^{\si,\sip}\!(\Lambda)
    = {1 \over {4\pi}}{{v(\sia)- v(\sipa)}\over {\sia -\sipa}}.}

Comparing \delflat\ to \invflat, we note that, because of the
boundary conditions at large distances, 
an extra term appeared on the right hand side of \invflat. This
term is the source of the conformal anomaly; in the string equations of \pol,
this anomaly is cancelled by the explicitly conformal non-invariant terms in
the action. The above equation will play an important role in the calculations
that follow: In applying the fundamental equation (2.38) to special
cases, one needs an explicit expression for the variation of the
propagator under the change of the cutoff; namely the term $\delta_{\Lambda}
\Delta^{\si,\sip}$ in the above equation. This equation therefore provides
 the needed explicit expression. Another point that needs to be clarified
is the dependence of $\Delta^{\si,\sip}$ on the variables $\si$ and
$\sip$. We would like to impose two dimensional rotation and translation
invariance on the world sheet even in the presence of the cutoff. There is
no problem in imposing both of these invariances for a fixed cutoff,
 however,  when the cutoff is changed infinitesimally
from this fixed value, its variation
 is given by \invflat\ and it is clearly no longer translation and rotation
invariant. This is the consequence of the
translation and rotation non-invariant long
distance boundary condition imposed in determining the cutoff variation.

The generators of conformal transformations, in the form they are expressed
in \spec, are not covariant under general coordinate transformations \diff.
They can be cast into a covariant form by writing
\eqn\specc{F^{\mu\si}_{v}(X)= v(\sia)\partial_{+}X^{\mu\si}
    +\mintop v(\sipa) f^{\mu\si}_{\sip}(X).}
The function $f$ is introduced to make $F$ transform as a vector in the
indices $\mu\si$; note that the subscript $\sip$ in $f^{\mu\si}_{\sip}$
is not a tensor index.  A further 
constraint on $f$ comes from demanding that $\delta_{v}$ satisfy the Virasoro
algebra. We will specify this function later on when we discuss concrete
examples.

We are now ready to write down the fundamental string field equation; it is
obtained by demanding  the invariance of the partition function  under
the conformal variation \var:
\eqn\deltaZ{\delta Z(X,\Lambda)=0.}
We note that this is not merely a requirement of covariance but one of 
invariance. In this respect, $F^{\mu\si}_{v}$ acts like a Killing vector
that generates the conformal symmetry. Carrying out the operations represented
by $\delta$ on the right hand side of \Zfull\ gives
\eqn\prel{\left.\exp\left(S_{I}(X,P)\right) {\cal H}(X,P)
    \exp\left({1 \over 2}J_{\mu\si}G^{\mu\si,\nu\sip}
    J_{\nu\sip}\right)\right|_{J=0}=0,}
where ${\cal H}(X,P)$ will be defined shortly. From this equation, 
it is tempting to conclude that
\eqn\wron{{\cal H}(X,P)=0.}
However, this conclusion is not correct; eq. \wron\ is too strong as it
stands.
This is because of the existence of an identity of the form
\eqn\ident{\left.\exp \left(S_{I}(X,P)\right) 
    K^{\mu\si}(X,P)\left({
    \delta S_{I} \over \delta P^{\mu\si}}+ G_{\mu\si,\nu\sip}
    P^{\nu\sip}\right) \exp \left({1 \over 2}J_{\alpha\sipp}
    G^{\alpha\sipp,\beta\sippp}J_{\beta\sippp}\right)\right|_{J=0}=0,}
where $K$ satisfies
\eqn\div{{\delta K^{\mu\si}\over \delta P^{\mu\si}}=0,}
but is otherwise an arbitrary function of $X$ and $P$. This identity,
easy to verify directly, can be understood as follows. Reversing the
steps leading from \newZ\ to \Zfull, one can get rid of the
operator $P$ and write the above identity as an integral over the
variable $\xi$. The identity is then satisfied by virtue of the
integrand being a total derivative. A total derivative corresponds
to an infinitesimal change of variable in the integral in \newZ;
therefore \ident\ is equivalent to the invariance of \newZ\ under
such a change of variable. Eq. \div\ expresses the restriction that the
Jacobian of this transformation is unity so as to leave the action
unchanged, and \wron\ amounts to deducing the vanishing of the integrand
from the vanishing of an integral and it is therefore too strong; the
correct equation should be
\eqn\corre{{\cal H}=
    K^{\mu\si}\left({\delta S_{I}\over \delta P^{\mu\si}}+ G_{\mu\si,
    \nu\sip} P^{\nu\sip}\right),}
and \prel\ is satisfied by virtue of \ident.

The function ${\cal H}$ can be determined by carrying out the variations
indicated in \deltaZ; the result is
\eqn\equation{\eqalign{& \left(F^{\mu\si}_{v}{\delta \over \delta
    X^{\mu\si}}+P^{\nu\sip} {\delta F^{\mu\si}_{v}\over \delta X^{
    \nu\sip}}{\delta \over \delta P^{\mu\si}}+\delta_{\Lambda} \right)
    \left( S(X)+S_I(X,P)-{1 \over 2}\Tr \log G(X)\right)\cr 
    &\quad\null-{1 \over 2}
    \left(F^{\alpha\sipp}_{v}{\delta G^{\mu\si,\nu\sip}\over\delta X^{
    \alpha\sipp}}-{\delta F^{\mu\si}_{v} \over \delta X^{\alpha\sipp}}
    G^{\alpha\sipp,\nu\sip}-{\delta F_v^{\nu\sip}\over \delta X^{\alpha
    \sipp}}G^{\mu\si,\alpha\sipp}+\delta_{\Lambda}G^{\mu\si,\nu\sip}
    \right)\cr &\quad\times\left( {\delta^{2} S_{I}(X,P) 
    \over \delta P^{\mu\si}
    \delta P^{\nu\sip}}+ {\delta S_{I} \over \delta P^{\mu\si}}
    {\delta S_{I} \over \delta P^{\nu\sip}} \right)= 
    K^{\mu\si}\left( {\delta S_{I} \over \delta
    P^{\mu\si}}+G_{\mu\si,\nu\sip} P^{\nu\sip}\right). \cr }}

Eq. \equation\ is our version of renormalization group equations for the string 
action $S$. As it stands, it has two unusual features:
\item{a)} It is an equation in two variables $X$ and $P$, whereas the 
standard renormalization group equations are in a single variable, the
background field $X$.
\item{b)} It contains a function $K^{\mu\si}(X,P)$, arbitrary except for the
constraint given by \div.
\bigskip

We will now show that these two seeming defects cancel each other; it is 
possible to convert \equation\ into an equation in a single variable
$X$ by taking advantage of the arbitrariness of the function $K$. To see
this, imagine expanding $S_{I}$, ${\cal M}$ and $K$ in a power series in
the variable $P$. By equating different powers of $P$ on both sides of the
equation, we obtain an infinite set of equations, each in the single
variable $X$. Let us now focus on the equation zeroth order in $P$. Since
the right hand side of \equation\ starts with a linear term in $P$,
this equation receives no contribution from $K$. This follows from the
fact that $S_{R}$, ${\cal M}$ and therefore $S_{I}$ all start at least
quadratically in the expansion in powers of $P$. We therefore have an
equation in the single variable $X$ and
free of the ambiguity coming from $K$:
\eqn\basic{E_{G}+ E_{\cal M}=0,}
where,
\eqn\basicc{E_{G}=\left( F^{\mu\si}_{v}{\delta \over \delta X^{\mu\si}}
    +\delta_{\Lambda}\right) \left(S- {1 \over 2}\Tr \log G\right),}
and
\eqn\basiccc{\eqalign{E_{\cal M}= &{1 \over 2}\left(- F^{\alpha\sipp}_{v}
    {\delta G^{\mu\si,\nu\sip}
    \over \delta X^{\alpha\sipp}}+ {\delta F^{\mu\si}_{v} \over \delta X^{
    \alpha\sipp}} G^{\alpha\sipp, \nu\sip} + {\delta F_v^{\nu\si} \over
    \delta X^{\alpha\sipp}} G^{\mu\si, \alpha\sipp}- \delta_{\Lambda}
    G^{\mu\si,\nu\sip} \right)\cr
    &\times \left[ {\delta^{2}{\cal M} \over 
    \delta P^{\mu\si}\delta P^{\nu\sip}}
    + {\delta {\cal M}\over \delta P^{\mu\si}}{\delta {\cal M} \over
    \delta P^{\nu\sip}}\right]_{P=0}=0.\cr}}

This equation is the fundamental result of the section. In the next two
sections, it will provide our starting point for the derivation of the
string field equations. 

We end this section with a couple of
comments:
\item{a)} Eq. \basic\ is a a one loop result, which one can verify either
by counting powers of $\hbar$ or more simply from the appearance of the
``Tr log'' type terms. It should therefore agree with the standard
treatment \callan\ for a renormalizable action $S$. We will make this
comparison in the next section.
\item{b)} This observation leads to an apparent paradox: No 
approximation has been made in deriving \basic, yet it is clearly a
one loop result. We have to conclude that the one loop result leads to
exact string field equations. 
\item{c)} There are of course an additional infinite number of
equations coming from higher powers of $P$. These
equations can then be used to determine the unknown function
$K^{\mu\si}(X,P)$. One can then extract  relations not involving
$K$ by using the constraint \div. These appear to come from two or more
loops. We do not know whether these equations are redundant,
 or whether they contain additional information, which would then supplement
the one loop result but not change it.
\item{d)} Eq. \basic\ contains two unknown functions: The Killing vector
$F^{\mu\si}_{v}$ and implicitly, the connection $\Gamma^{\gamma\sipp}
_{\alpha\si,\beta\sip}$ (See \G). They have to be expressed in terms
of the fields that appear in $S$. This will be done in the following
sections.


\newsec{The Tachyon and the Massless Level}

In this section, we will apply the formalism developed in the last section
to the two lowest levels of a closed, unoriented bosonic string, the
tachyon and the massless level. Ideally, one would like to start with
$S$ as an arbitrary functional of $X^{\mu\si}$ and try to solve
\basic\ in all its generality. However, this direct approach
  seems hopelessly complicated
and not particularly useful. Instead, the problem is made tractable 
by expanding $S$ in powers of the $\si$ derivatives of $X^{\mu\si}$.
We will call this the quasi-local expansion. This expansion is quite
natural from the point of two dimensional field theory on the world
sheet and it has been the basis of most of the work done on this
subject. From the string point of view, it is an expansion in the
level number, two derivatives in $\si$ corresponding to an increase
of one unit in level number. We  note that as a consequence of
two-dimensional rotation invariance on the world sheet, which we shall
always assume, there is always an equal number of derivatives with
respect to $\sia$ and $\sis$. It is then convenient to introduce a
parameter ``$b$'' to keep track of the expansion: The field representing
the nth level of the string will be multiplied by $b^{(n-1)}$. For example,
the tachyon has coefficient $b^{-1}$ and the massless level is 
independent of $b$. The first two terms in
the quasi-local expansion of $S$ are then given by
\eqn\ql{ S= b^{-1}S^{(-1)}+ S^{(0)}+ \cdots =
    \minto \left(b^{-1} \Phi(X(\si)) +\tilde{g}_{\mu\nu}
    (X(\si)) \partial_{+}X^{\mu\si} \partial_{-}X^{\nu\si} \right).}

Here and in the sequel, we have adopted the following notation:
The superscripts ($-$1), (0), etc., 
refer to terms in $S$ proportional
to the corresponding powers of $b$.
Expressions like $\Phi(X(\si))$ denote
local functions of the coordinate $X(\si)$, whereas expressions such as
$F^{\mu\si}(X)$ denote functionals in the same coordinate. Also, we
should make clear that the parameter ``$b$'' is merely a bookkeeping device
and can be set equal to one at the end of the calculation.

In the
same spirit, the coordinate transformations \diff\ have a quasi-local
expansion:
\eqn\qlt{ F^{\mu\si}(X)= f^{\mu}(X(\si)) + b f^{\mu}{}_{\nu\lambda}
    (X(\si)) \partial_{+}X^{\nu\si}\partial_{-}X^{\lambda\si}+ 
    bf^\mu{}_\nu \dpm X^{\nu\si}(X(\si)) +\cdots.}

The first term is the local coordinate transformation associated with
gravity; terms with increasing powers of $b$ contain higher derivatives
of $\si$ and become increasingly non-local. In this section, we will only be
concerned with invariance under local transformations represented by the first
term in \qlt. However, the general
strategy, pursued in the next section, is  to determine
$F^{\mu\si}_{v}$, the generator of the conformal transformation 
(see \specc), and the connection $\Gamma$ as a power series in $b$ so as 
to achieve covariance under both local and
non-local transformations. To zeroth order in $b$, $F$,
the generator of the conformal transformations, is given by the first 
term in \specc; the function $f^{\mu\si}_{v}$ is at least of first order
in $b$.

To simplify the exposition, we have so far neglected the cutoff dependence 
in $S$ (see \action). With the cutoff restored, the second term in \ql\
should read
\eqn\qlc{S^{(0)}= \mintt \tilde{g}_{\mu\si,\nu\sip}\partial_{\sia}X^{\mu\si}
    \partial_{\sips}X^{\nu\sip},}
where  $\tilde{g}$ is given by
\eqn\qlcg{\tilde{g}_{\mu\si,\nu\sip}= \eta_{\mu\nu}\diraccutoff(\si,\sip)
    + \tilde{h}_{\mu\nu}(X(\si)) \dirac(\si -\sip).}
$\diraccutoff(\si,\sip)$ is defined in \invc\ and
$\tilde{h}$ is a cutoff independent local function of $X(\si)$.

We have now to determine the connection $\Gamma$ and the generator
of conformal transformations $F^{\mu\si}_{v}$ to zeroth order in $b$.
We have already observed above that $F$ is given by \spec\
to zeroth order, since $f$ in \specc\ is already first order in $b$.
As for the connection, it will be derived from a metric that transforms
correctly under local transformations.
The standard choice for the metric made in sigma model
calculations, which we shall adopt, is the symmetric part of $\tilde{g}$
in \qlc:
\eqn\metric{g_{\mu\si,\nu\sip}=\ha \left(\tilde{g}_{\mu\si,\nu\sip} +
\tilde{g}_{\nu\sip,\mu\si}\right)= \eta_{\mu\nu}\diraccutoff(\si,\sip)
    + h_{\mu\nu}(X(\si)) \dirac(\si - \sip),}
where,
\eqn\metricc{h_{\mu\nu}= \ha \left(\tilde{h}_{\mu\nu}+
    \tilde{h}_{\nu\mu}\right).}

The connection, to zeroth order in $b$, is given in terms of the metric
by the standard formula:
\eqn\gamm{\Gamma^{\mu\si}_{\alpha\sip,\beta\sipp}= \ha
    g^{\mu\si,\lambda\sippp}\left({\delta g_{\lambda\sippp,\beta\sipp} \over
    \delta X^{\alpha\sip}}+ {\delta g_{\alpha\sip,\lambda\sippp} \over
    \delta X^{\beta\sipp}}- {\delta g_{\alpha\sip,\beta\sipp} \over
    \delta X^{\lambda\sippp}}\right).}

With these preliminaries out of the way, we are ready to write down the field
equation for the tachyon field. This we do by extracting terms lowest order
in $b$, proportional to $b^{-1}$, from \basic. Notice that only $E_{G}$
contributes. The equation reduces to
the following simple form
\eqn\eqfordil{\left(\minto v(\sia)\partial_+ X^{\mu\si}
    {\delta \over \delta X^{\mu\si}}+\delta_{\Lambda}\right)
    \left(S^{(-1)}_{\nt}[X]- \ha \Tr \log G^{(-1)}\right)=0.}

The first term on the left is easy to calculate:
\eqn\first{\minto v(\sia)\del_+ X^{\mu\si}{\delta \over \delta
X^{\mu\si}}\left(S^{(-1)}_{\nt}[X]\right)
 = -\minto v'(\sia)\Phi(X(\si)).}

In calculating the contribution of the second term, we take advantage of the
following simplifications: We are going to drop all the non-local terms
that can arise in the expansion of this term. Such non-local terms are in
general present since the action $S$ that satisfies \equation\
 is not necessarily one particle irreducible, and we
wish to extract the one particle irreducible part that is local.
Another
simplification follows from the fact that the result clearly is going
to be a covariant
Klein-Gordon equation for the tachyon field $\Phi$ in the background
metric $g$. We can then first linearize this equation by expanding to
first order in $h_{\mu\nu}$ of \metric\ around the flat background,
and then covariantize the result to arrive at the full answer in an
arbitrary background. This will be our strategy in the rest of the paper;
only the linear part of the field equations will be computed in the
presence of a  flat background, and the result will be 
generalized to  a non-trivial background, making use of the powerful
restrictions resulting from covariance. 

Eq. \eqfordil\ is linearized by setting
\eqn\Gflath{G_{\mu\si,\nu\sip}(X) = 2  \Delta_{\mu\si,\nu\sip}
        +H_{\mu\si,\nu\sip}(X),}
and by expanding the ``Tr log'' to first order in  $H$:
\eqn\tlG{\Tr \log G 
        \cong {1\over 2}\Delta^{\mu\si,\nu\sip}H_{\mu\si,\nu\sip}.}

The linear part of $H$, to order $b^{-1}$, is given by
\eqn\second{H^{(-1)}_{\mu\si,\nu\sip}\cong {{\dirac S_{\nt}[X]}\over
        {\delta X^{\mu\si} \delta X^{\nu\sip}}}\cong \dirac (\si-\sip)
        \partial_\mu \partial_\nu \Phi(X(\si)) .}

In calculating the left hand side of \eqfordil, the following identity
proves useful:
\eqn\lemm{\eqalign{&\Delta^{\mu\si,\nu\sip}\mintopp v(\sippa)
    \partial_{+}X^{\lambda\sipp} {\delta \over \delta X^{\lambda\sipp}}
    \left({\delta^{2} S \over \delta X^{\mu\si} \delta X^{\nu\sip}}\right)=
    \Delta^{\mu\si,\nu\sip}{\delta^{2}\over\delta X^{\mu\si}\delta
    X^{\nu\sip}}\cr &\qquad\qquad\times\left(\mintopp 
    v(\sippa)\partial_{+} X^{\lambda\sipp}
    {\delta S \over \delta X^{\lambda\sipp}}\right) - {\delta^{2} S \over
    \delta X^{\mu\si} \delta X^{\nu\sip}} \left(v(\sia)\partial_{\sia}+
    v(\sipa)\partial_{\sipa}\right) \Delta^{\mu\si,\nu\sip}.\cr}}

The second term on the right hand side  can be calculated
making use of \invflat, leading to the equation
\eqn\tach{\minto v'(\sia) \left[ -\Phi(X(\si)) + {1\over 16\pi}
    \partial^{\mu} \partial_{\mu} \Phi(X(\si))+ {1\over 4}\Delta^{(0)}
    (\Lambda) \partial^\mu \partial_\mu \Phi(X(\si)) \right] = 0, }
where $\Delta^{(0)}(\Lambda)$ is the propagator
$\Delta^{\si,\sip}(\Lambda)$, evaluated at
$\si=\sip$. By translation invariance, it is independent of $\si$.
 Since $\Phi$ is only a function of $X^{\mu\si}$ and not
of its derivatives with respect to $\si$, it follows that
\eqn\tachy{\left({1 \over 16 \pi}+ {1 \over 4} \Delta^{(0)}(\Lambda)
    \right) \partial^{\mu}\partial_{\mu}\Phi(X(\si)) - \Phi(X(\si))=0.}

The term $\Delta^{(0)}(\Lambda)$ is cutoff dependent and it blows up as
$\Lambda \rightarrow \infty$. This cutoff dependent term can be eliminated
by explicitly introducing the slope parameter which we have suppressed and
by renormalizing it. The same cutoff dependent term is encountered in the
equations for the higher levels and it is again eliminated by the same
slope renormalization. Finally, \tachy\ can easily be generalized to an 
arbitrary background by using the metric
given by \metric\ and casting it into a covariant form.

The next step is to derive the field equation for $\tilde{g}$,
which includes both the metric (\metric), and the antisymmetric tensor
\eqn\antis{B_{\mu\nu}= \ha (\tilde{h}_{\mu\nu}- \tilde{h}_{\nu\mu}).}

To do this, we have to extract zeroth order terms in $b$ from \basic.
It is useful to distinguish between the two terms $E_{G}$ and $E_{\cal M}$,
the former  coming from the variation
of the $\Tr \log G$, and the latter coming from the variation of ${\cal M}$.
 The reason for this distinction is that the
$E_{G}$ is cutoff independent, whereas $E_{\cal M}$ is
proportional to a cutoff dependent factor. We will argue later that these
two terms must vanish separately, yielding two separate equations.
The first of these will be the equation for the metric $g$ and the 
antisymmetric tensor $B$; the second will provide the equation for the
dilaton. Our strategy is again to expand around the flat background to
first order in $h$ and $B$, and use covariance to arrive at the full
answer. We make use of \tlG\ to
calculate  $\Tr \log G$,  extracting the linear piece in $\tilde{h}$,
and using \invflat, we find
\eqn\varH{\eqalign{E_{G}^{(0)}=&\left.\left(\minto v(\sia)\del_+ X^{\mu\si}
         {\delta \over {\delta X^{\mu\si}}} + \delta_\Lambda \right)
       \left (H^{(0)}_{\mu\sip,\nu\sip}
 \Delta^{\mu\sip,\nu\sip}(\Lambda) \right)\right|_{{\rm lin}}=\cr
        =& -{1\over 4\pi} \mintt {{v(\sia)-v(\sipa)}\over{\sia-\sipa}}
 \left(\!\left. {{\dirac S^{(0)}_{\nt}[X]}\over
 {\delta X^{\mu\si}\delta X^{\mu\sip}}}\right|_{\rm lin}\!\!\!
        +2 \dirac(\si-\sip)\Gamma^\lambda_{\mu\mu}(X(\si))\dpm
        X^{\lambda\si}\! \right). \cr }}

Here, the subscript ``lin'' refers to  terms linear in $\tilde{h}$.
The $\Gamma$ that appears on the right hand side is the linearized form
of the connection \gamm:
\eqn\connect{\Gamma^{\mu\si}_{\alpha\sip,\beta\sipp}\cong 
    {\tilde \delta}^2_\Lambda(\si,\sip) \dirac(\sip - \sipp)
    \Gamma^{\mu}_{\alpha\beta}(X(\sip)),} 
where,
\eqn\connectt{ \Gamma^{\mu}_{\alpha\beta}(X(\si)) \cong \ha
    \eta^{\mu\nu} \Big(\partial_{\alpha} h_{\nu\beta}(X(\si))
    + \partial_{\beta} h_{\alpha\nu}(X(\si)) - \partial_{\nu}
    h_{\alpha\beta}(X(\si)) \Big).}
The 
functional derivative of $S^{(0)}_{\nt}$ can be calculated from \ql,
and repeating the steps that led to \tachy\ gives the following
field equation:
\eqn\vplus{\bbox \tilde{h}_{\nu\lambda}
    -\partial_\nu \partial_\mu \tilde{h}_{\mu\lambda}    
    + \partial_\mu \partial_\lambda \tilde{h}_{\mu\nu}
    - 2\partial_\lambda \Gamma^\nu_{\mu\mu}=0,}
where $\bbox = \partial_\mu\partial^\mu$. Here, since we use a flat metric
to raise and lower indices, there is no real distinction between upper
and lower indices. This will be understood whenever we have repeated upper
or lower indices.

The equation above came from the conformal transformations in the
variable $\sia$. The other set of conformal transformations in the
variable $\sis$ result in an additional equation:
\eqn\vminus{\bbox \tilde{h}_{\nu\lambda}-
\partial_\lambda \partial_\mu \tilde{h}_{\nu\mu}
    + \partial_\nu \partial_\mu \tilde{h}_{\lambda\mu}
    - 2\partial_{\nu} \Gamma^\lambda_{\mu\mu}=0.}

It is now convenient to combine these two equations and rewrite them
in terms of $h$ and the antisymmetric tensor $B$. Interestingly, we find
that, without any reference to \connectt, these equations fix the contracted
connection $\Gamma$ up to an
arbitrary scalar field $\phi$, which we shall identify with the dilaton
field:
\eqn\subtract{\Gamma^\lambda_{\mu\mu}= \partial_\mu h_{\mu\lambda}
            +\partial_\lambda \phi,}
and we arrive at the following equations for $h$ and $B$:
\eqn\pauli{\bbox h_{\nu\lambda}-\partial_\nu \partial_\mu h_{\mu\lambda}
        -\partial_\lambda \partial_\mu h_{\mu\nu}-2\partial_\nu \partial
        _\lambda \phi =0,}
\eqn\bloch{\bbox B_{\nu\lambda}-\partial_\nu \partial_\mu B_{\mu\lambda}
        +\partial_\lambda \partial_\mu B_{\mu\nu}=0.}
Comparing with \connectt\ determines $\phi$:
\eqn\Thetta{\phi = -\ha h_{\mu\mu},}
and substituting this result back into \pauli\ gives the standard
equations of gravity without source in linearized form. The equation
for the antisymmetric tensor $B_{\mu\nu}$ is the linearized version of the
standard result of \callan.

Up to this point, we have not taken into account $E_{\cal M}$.
Let us first calculate the ${\cal M}$ dependent factor in \basiccc.
From equation \M, we find that
\eqn\dilequat{\eqalign{&\left. {{\dirac {\cal M}}\over{\delta P^{\mu\si} 
            \delta P^{\nu\sip}}}\right|_{P=0}
    = -{1\over3}\left({{\delta \Gamma^{\lambda\sipp}_{\lambda\sipp,\mu\si}}
            \over{\delta X^{\nu\sip}}} + {{\delta
            \Gamma^{\lambda\sipp}_{\lambda\sipp,\nu\sip}}
            \over{\delta X^{\mu\si}}}+
            {{\delta \Gamma^{\lambda\sipp}_{\mu\si,\nu\sip}}
            \over{\delta X^{\lambda\sipp}}} \right) \cr
     &\qquad= -{1 \over 3} \Delta^{(0)}(\Lambda) \dirac(\si -\sip) \Big(
    \partial_{\nu} \Gamma^{\lambda}_{\lambda\mu}(X(\si)) + \partial_{\mu}
    \Gamma^{\lambda}_{\lambda\nu}(X(\si)) + \partial_{\lambda}\Gamma^{
    \lambda}_{\mu\nu}(X(\si))\Big).\cr}}

The term which is  quadratic in ${\cal M}$ will not contribute since it
is non-linear (and also non-local). 
 To the order we are considering, the
 first factor on the right in \basiccc\ can be evaluated by setting
$G^{\alpha\si,\beta\sip}$ equal to $\Delta^{\alpha\si,\beta\sip}$, with the
result,
\eqn\thisgives{\eqalign{E_{\cal M}^{(0)}&=
    \ha \left( v(\sia)\partial_\sia \Delta^{\mu\si,
    \nu\sip} + v(\sipa)\partial_\sipa \Delta^{\mu\si,\nu\sip}
    - \delta_{\Lambda} \Delta^{\mu\si,\nu\sip} \right)
    \left. {{\dirac {\cal M}}\over{\delta P^{\mu\si} 
    \delta P^{\nu\sip}}}\right|_{P=0}\cr
    &=\null  -{1 \over 24 \pi} \Delta^{(0)}(\Lambda) \eta^{\mu\nu}\minto 
    v'(\sia)\left(2\partial_\nu
    \Gamma^\lambda_{\lambda\mu}(X(\si))+\partial_\lambda \Gamma^\lambda
    _{\mu\nu}(X(\si)) \right). \cr}}

We see that unlike the cutoff independent $E_{G}$, $E_{\cal M}$ is proportional
to the cutoff dependent factor $\Delta^{(0)}(\Lambda)$. It easy to show that
this term cannot be eliminated by the addition of any local counterterm to
$S$, and therefore, it must be set equal to zero by itself. This gives us
the additional equation
\eqn\equatdil{2\partial_\mu\Gamma^\lambda_{\lambda\mu}+
    \partial_\lambda \Gamma^\lambda_{\mu\mu}= 
    \ha \bbox h_{\mu\mu}+ \partial_{\mu} \partial_{\nu} h_{\mu\nu}=0,}
and combining this with \pauli\ and \Thetta, we find that
\eqn\planck{\bbox h_{\mu\mu}= 0, \qquad
    \partial_{\mu} \partial_{\nu}h_{\mu\nu}=0,}
and
\eqn\electron{\bbox h_{\mu\nu}- \partial_{\mu} \partial_{\lambda}h_{\lambda\nu}
-\partial_{\nu}\partial_{\lambda}h_{\lambda\mu}- \partial_{\mu}\partial_{\nu}
h_{\lambda\lambda}= 0.}

Equations \planck\ and \electron\ describe the coupled graviton-dilaton system
in the linear approximation. To see this, we note that these equations are
invariant only under coordinate transformations of unit determinant, which,
linearized, results in invariance under gauge transformations
\eqn\shift{h_{\mu\nu}\rightarrow h_{\mu\nu}+ \partial_{\mu}\kappa_{\nu}
+ \partial_{\nu}\kappa_{\mu},}
with the important restriction that $\partial_{\mu}\kappa^{\mu}=0$.
 This restriction to unit determinant was
explained in section 2 in the paragraph following  \ggamm. As a consequence,
the trace of $h$, $h_{\mu\mu}$, which can be gauged away if there is invariance
under unrestricted coordinate transformations, can no longer be eliminated and
becomes a dynamical degree of freedom. Up to normalization, we identify it 
with the dilaton field $\phi$. The natural candidate for the graviton field
 is the traceless component of $h$:
\eqn\traceless{\bar{h}_{\mu\nu}\equiv h_{\mu\nu}- {1 \over D}\eta_{\mu\nu}
h_{\lambda\lambda},}
where D is the dimension of space. We identify $\bar{h}$ with the graviton
 field in the gauge where the metric has unit determinant and as a 
consequence, the graviton field is traceless. $\bar{h}$ also satisfies
\electron, which is the correct equation for the graviton coupled to the
dilaton in this gauge. We would like to point out the difference between our
treatment of the dilaton and the standard approach. In the standard treatment,
in addition to $X^{\mu\si}$, the dilaton field $\phi$ is introduced in the
 action from the beginning, and the theory is regularized by going from 2
to $2+\epsilon$ dimensions on the world sheet. We stay with a two dimensional
world sheet, regularize only the free propagator (see \action), and the 
dilaton field is identified with the log of the determinant of the metric.
This identification is only possible because the full coordinate invariance
is broken down to transformations of unit determinant.

We end this section with a few observations:
\item{a)} So far, we have worked out the coupled system of the graviton,
 dilaton and the antisymmetric tensor only in the linear approximation. As
stressed earlier, the full dependence on the graviton field follows from
covariance. However, we have not calculated the higher order contributions
in the dilaton field $\phi$ and the antisymmetric tensor field $B_{\mu\nu}$.
It would be interesting to compare these to the results of \callan, although
such a comparison is plagued with ambiguities due to possible field
redefinitions involving the dilaton field. It is also not clear that we
should even consider the antisymmetric tensor: Our approach works only
for the left-right symmetric string models and the antisymmetric tensor 
decouples in that case.
\item{b)} In the presence of the cutoff, the coordinate transformation, given
by \qlt, has to be modified to preserve the invariance of the action. In the
linear approximation, the modification is
\eqn\modi{F^{\mu\si}(X) \cong \mintop {\tilde\delta}^2_
    \Lambda(\si,\sip)
    f^{\mu}(X(\sip))+\cdots.}
Although they are not needed in this paper, the non-linear corrections to \qlt\
can, in principle, be worked out.
\item{c)} There is an ambiguity in the expression for the connection given by
\connectt, which is the standard result of differential geometry
derived from the metric. However, since we insist on invariance under
transformations with unit determinant, we are free to modify the metric by,
for example
\eqn\bragg{ g_{\mu\nu} \longrightarrow g_{\mu\nu}(\det g)^k,}
where $k$ is an arbitrary constant. The modified metric leads to a modified
connection, and to a new set of equations. These equations are not, however,
physically different from \planck\ and \electron; they correspond to field
redefinitions involving the dilaton field mentioned above. This becomes
clear by noticing that the dilaton field can be taken to be the log of the
determinant of $g$; then \bragg\ is a dressing of 
the metric by the dilaton field.
\item{d)} It is of some interest to find out what would have happened, if we
had carried out a non-covariant calculation. This means setting the
connection $\Gamma$ equal to zero throughout, and referring to the equations
\vplus\ and \vminus, it amounts to choosing the gauge
\eqn\choos{\Gamma^{\lambda}_{\mu\mu}\cong \partial_{\mu}h_{\mu\lambda}
    - \ha \partial_{\lambda} h_{\mu\mu}=0.}
Therefore, the equation for the graviton comes out gauge fixed, but
otherwise correct. What is missing is \planck, the equation for the dilaton
field. This is because the equation of motion for the dilaton comes entirely
from ${\cal M}$, and with connection equal to zero, ${\cal M}$ is also zero.


\newsec{The First Massive Level - Non-Covariant Approach}

In this section, we shall investigate the first massive state, using the
tools developed in section 2. The particular question we would like to address 
 is whether the spectrum of states that follows from the linear (free) part of
the equations of motion we are going to derive is
consistent with the known spectrum of the first massive level of the string.
This is clearly a necessary test
 any successful candidate for string field equations
must pass. Of course, in addition, the non-linear part of the equations should
reproduce the interactions of the string theory. We will not address the
 question of interactions here, apart from observing that the stringent
requirements of covariance we are going to impose probably fix the interaction
uniquely.

The field equations will again follow from \basic, given $F$ (\specc) 
and the connection $\Gamma$  to first order in $b$. For the sake of
comparison with the non-covariant renormalization group approach of \pol,
we will first carry out a calculation with vanishing connection and $F$ given
by \spec.
Comparing the resulting
physical states to those of the string, we will find that there are too many
of them. In the next section
the calculation is done covariantly:
We start with $\Gamma$ and $F$ derived from a metric,
suitably defined  so as to
satisfy invariance under coordinate transformations \diff\ and \deter\ to
first order in $b$. The resulting set of states appear to be consistent with
those of the left-right symmetric string model.
  We conclude that only the covariant
approach yields equations powerful enough to produce the spectrum of at least
the left-right symmetric  string
theory; the equations resulting from the non-covariant approach turn out to be
too weak.

The starting point is the first massive level, written out in full generality:
\eqn\general{\eqalign{S^{(1)}= \minto &\left( e^{(1)}_{\m1\m2,\n1\n2} \dpmu1
        \dpmu2 \dmnu1 \dmnu2 + e^{(1)}_{\m1 \m2 \m3} \dpmmu1 \dpmu2 \dmmu3
        \right. \cr
        &+ e^{(2)}_{\m1 \m2 \m3} \dppmu1 \dmmu2 \dmmu3 + e^{(3)}_{\m1 \m2 \m3}
        \dmmmu1 \dpmu2 \dpmu3 \cr &+ e^{(1)}_{\m1 \m2} \dpmmu1 \dpmmu2 
         +e^{(2)}_{\m1 \m2} \dppmmu1 \dmmu2 \cr
        &+ \left. e^{(3)}_{\m1 \m2} \dmmpmu1 
        \dpmu2 +e^{(4)}_{\m1 \m2} \dppmu1 \dmmmu2 \right), \cr }}
where the $e$'s in this expression are local 
functions of the field $X^{\mu\si}$.
Here and in many of the equations that follow, we have also 
simplified writing by replacing, for
example, $X^{\mu_{1}\si}$ by $X^{\mu_{1}}$.

Eq. \general\ is highly redundant because of the existence of linear gauges.
These result from the possibility of adding zero to \general\ by adding a
total derivative in $\sia$ or in $\sis$ to the integrand. Such a possibility
already exists for the zero mass level; adding
\eqn\zer{0= \minto \left(\partial_{+}\left(\partial_{-}X^{\mu\si}
    \Lambda_{\mu}(X(\si))\right) - \partial_{-}\left(\partial_{+}X^{\mu\si}
    \Lambda_{\mu}(X(\si)) \right)\right)}
to \ql\ amounts to the well-known gauge transformation of the antisymmetric
tensor $B$:
\eqn\gaugeB{ B_{\mu\nu} \rightarrow B_{\mu\nu} + \partial_{\mu}\Lambda_{\nu}
- \partial_{\nu}\Lambda_{\mu}.}

For the first massive level, the situation is more complicated; there are six
distinct linear gauge transformations. These are discussed in Appendix A, where
it is also shown that, making use of these gauges, all but three of the fields
appearing in \general\ can be eliminated. The resulting linear gauge fixed
form of $S^{(1)}$ reads
\eqn\bytaking{\eqalign{S^{(1)}=\minto &\left( e_{\m1\m2,\n1\n2}\dpmu1\dpmu2
        \dmnu1\dmnu2 \right. \cr
        & \left.\null+e_{\m1\m2\m3}\dpmmu1\dpmu2\dmmu3+
        e_{\m1\m2}\dpmmu1\dpmmu2\right).\cr}}

It is also shown in Appendix A that this form of $S^{(1)}$
is in fact completely gauge fixed;  in contrast to the massless level,
there are no linear gauge transformations left
of the form \gaugeB\ that map it
into itself. It is now easy to carry out the non-covariant calculation by
substituting $S$ given by \bytaking\ in \basic, and setting $\Gamma=0$ and
$F$ to the value given by \spec. The resulting equation is
\eqn\again{ \left( \mintopp v(\sippa)\partial_+ X^{\lambda\sipp} 
        {\delta \over {\delta X^{\lambda\sipp}}} + \delta_\Lambda \right)
        \left(S^{(1)}-{1\over 4}
        \Delta^{\mu\si,\nu\sip}{{\dirac S^{(1)}} \over {\delta X^{\mu\si}
        \delta X^{\nu\sip}}}\right)=0.}

The second term of this equation can be evaluated with help of the
identities \lemm\ and \invflat:
\eqn\andagain{\eqalign{&\left( \mintopp v(\sippa)\partial_+ X^{\lambda\sipp}
        {\delta \over {\delta X^{\lambda\sipp}}} + \delta_\Lambda \right)
        \Delta^{\mu\si,\nu\sip}{{\dirac S^{(1)}} \over {\delta X^{\mu\si}
        \delta X^{\nu\sip}}}= \cr
        &= \Delta^{\mu\si,\nu\sip}\!{\dirac  \over {\delta
        X^{\mu\si}\delta X^{\nu\sip}}}\!\mintopp\! v(\sippa)
        \partial_\sippa X^{\lambda
        \sipp}\!{{\delta S^{(1)}}\over{\delta X^{\lambda\sipp}}} -{1\over4\pi}
        \!\mintt\! \!{{v(\sia)-v(\sipa)}
        \over{\sia-\sipa}}{{\dirac S^{(1)}} \over
        {\delta X^{\mu\si}\delta X^{\mu\sip}}}. \cr}}

Setting  $S^{(1)}=\minto U(X(\si))$ and using the above results gives
\eqn\oranges{\eqalign{\minto v'(\sia) U(X(\si))& -{1\over 4}
        \Delta^{\mu\si,\nu\sip}
        {\dirac \over {\delta X^{\mu\si} \delta X^{\nu\sip}}}\mintopp v'(\sippa)
        U(X(\sipp)) \cr &\qquad\qquad\null +{1\over16\pi }\mintt 
        {{v(\sia)-v(\sipa)}\over{\sia-\sipa}}
        {{\dirac S^{(1)}}\over{\delta X^{\mu\si}\delta X^{\nu\sip}}}=0. \cr}}

The last term in this equation can be evaluated after a 
tedious but straightforward calculation, with the result
\eqn\soso{\eqalign{&\mintt  {v(\sia)-v(\sipa) \over \sia -\sipa}
        {\dirac S^{(1)} \over
        \delta X^{\mu\si}\delta X^{\mu\sip}}
        =\minto v'(\sia)\cr &\qquad\times\left(\!( \bbox 
        e_{\m1\m2,\n1\n2}-2\partial_\m1
        \partial_\mu e_{\mu\m2,\n1\n2}+{1\over3}\partial_\m1\partial_\m2
        e_{\mu\mu,\n1\n2})\dpmu1\dpmu2\dmnu1\dmnu2 \right. \cr
        &\qquad\qquad ( \bbox e_{\m1\m2\m3}-4\partial_\mu e_{\mu\m2,\m1\m3}
        -\partial_\m2\partial_\mu e_{\m1\mu\m3}+{4\over3}\partial_\m2 e_{\mu
        \mu,\m1\m3})\dpmmu1\dpmu2\dmmu3 \cr
        &\qquad\qquad (\bbox e_{\m1\m2}-\partial_\mu e_{\m1\mu\m2}+{2\over3}
        e_{\mu\mu,\m1\m2})\dpmmu1\dpmmu2 \cr
        &\qquad\qquad (-2\partial_\mu e_{\mu\m1,\m2\m3}+{1\over3}\partial_\m1
        e_{\mu\mu,\m2\m3})\dppmu1\dmmu2\dmmu3 \cr
        &\qquad\qquad \left.(-\partial_\mu e_{\m1\mu\m2}+{2\over3}
        e_{\mu\mu,\m1\m2})\dppmmu1\dmmu2 \right). \cr}}

One has to take into account possible gauge invariance of the integral on the
right hand side of this equation. Because of the presence of
the factor $v'(\sia)$, the
gauges are generated by adding a total derivative with respect to $\sis$ only,
and as a result, there are only three of them, as opposed to six in the case of
\general. In writing down \soso, we have already eliminated all
redundant terms  and fixed the linear gauges completely.

Let us now evaluate the second term in \oranges. As opposed to \soso, which
is cutoff independent, here we encounter only cutoff dependent terms.
These terms
are proportional to $\Delta^{\si,\sip}(\Lambda)$ and its derivatives,
evaluated at $\si=\sip$. By rotation invariance on the world sheet, the number
of derivatives with respect to $\sia$ must match those with respect to $\sis$.
Defining
\eqn\deris{\eqalign{&\partial_{\sia}\partial_{\sis}
    \Delta^{\si,\sip}(\Lambda)|_{\si=\sip} 
    \equiv \Delta_{2}^{(0)}(\Lambda),\cr
    &\partial^{2}_{\sia}\partial^{2}_{\sis}\Delta^{\si,\sip}
    (\Lambda)|_{\si=\sip}
    \equiv \Delta_{4}^{(0)}(\Lambda), \cr}}
we have,
\eqn\sowhat{\eqalign{\Delta^{\mu\si,\nu\sip}&{\dirac \over \delta X^{\mu\si}
    \delta X^{\nu\sip}}\left.\mintopp  v'(\sippa) U(X(\sipp))
    \right|_{\rm sing}=  \cr 
    =& \minto v'(\sia) \left( \Delta^{(0)}(\Lambda) \bbox
    e_{\m1\m2,\n1\n2}\dpmu1\dpmu2\dmnu1\dmnu2\right. \cr
    &\left.\null+\bbox e_{\m1\m2\m3}\dpmmu1\dpmu2\dmmu3+
    \bbox e_{\m1\m2}\dpmmu1\dpmmu2
    \right) \cr &\left.\null +\Delta_{2}^{(0)}(\Lambda) \left(-8
    e_{\mu\m1,\mu\m2} +2 \partial_{\mu} e_{\mu\m1\m2}+ 2 \partial_{\m2}
    e_{\m1\mu\mu} - 4 \partial_{\mu}\partial_{\m2} e_{\mu\m1}\right)
    \dpmu1 \dmmu2\right. \cr
    &\left.\null + 2 \Delta_{4}^{(0)}(\Lambda) e_{\mu\mu}\right).}}

These singular terms can be eliminated by renormalization as follows: The
same slope renormalization that got rid of the cutoff dependent term in the
equation for the tachyon (see \tachy\ and the discussion that follows) also
eliminates the term proportional to $\Delta^{(0)}(\Lambda)$ here. The other
cutoff dependent terms are due to the contraction of two $X$'s in the same
vertex, and they can be taken care of by vertex renormalization. This amounts
to eliminating them by introducing local counterterms in $S$ of the form,
for example,
\eqn\counter{\Delta S = \const \times \Delta_{2}^{(0)}(\Lambda)
    \minto e_{\mu\m1,\mu\m2} \dpmu1 \dmmu2.}
 
In the operator formulation of the string theory, these divergent terms are 
eliminated by the operator normal ordering of the vertex.

After renormalization, one is left with the finite equations given by \soso.
They fall into two classes: Propagating equations of motion are (with the
unconventionally normalized mass squared given by 16 $\pi$)
\eqn\threeeq{\eqalign{\bbox e_{\m1\m2,\n1\n2}+16\pi  e_{\m1\m2,\n1\n2} &=0, \cr
        \bbox e_{\m1\m2\m3}+16\pi  e_{\m1\m2\m3}&=0, \cr
        \bbox e_{\m1\m2}+16\pi  e_{\m1\m2}&=0, \cr}}
plus constraints
\eqn\cconst{\eqalign{\partial_\mu e_{\mu\m1,\n1\n2}-{1\over6}\partial_\m1
            e_{\mu\mu,\n1\n2}&=0, \cr
            \partial_\mu e_{\m1\mu\m2}-{2\over3}e_{\mu\mu,\m1\m2}&=0, \cr}}
and also the constraints that come from $v'(\sis)$
\eqn\fromvp{\eqalign{\partial_\nu e_{\m1\m2,\nu\n2}-{1\over6}\partial_\n2 
        e_{\m1\m2,\nu\nu} &=0, \cr
        \partial_\mu e_{\m1\m2\mu}-{2\over3}e_{\m1\m2,\nu\nu}&=0. \cr}}

Comparing with the structure of the first massive level of the string (see
Appendix B), it is clear that the above constraints are too weak. For example,
in string theory, everything is expressible in terms of the analogue of
$e_{\m1\m2,\n1\n2}$, whereas here $e_{\m1\m2}$ and most of $e_{\m1\m2\m3}$
cannot be so expressed. Clearly, the latter fields are spurious should somehow
be eliminated. In the next section, we will see that the covariant
approach overcomes this problem.

\newsec{The First Massive Level - Covariant Approach} 
 
 
In this section, the field equations for the first massive level will be
rederived, this time imposing covariance under coordinate transformations
given by \qlt. When treating the massless levels,  covariance under only
the local transformations (first term in \qlt) was imposed; we now require,
in addition, covariance under transformations first order in $b$. We will
initially simplify the problem by starting with flat Minkowski metric, with
$\tilde{h}=0,$
in \qlcg, and with the action
\eqn\actionn{ S= X^{\mu\si}\Delta_{\mu\si,\nu\sip}(\Lambda) X^{\nu\sip}
    + b S^{(1)}=S^{(0)}+ b S^{(1)},}
where $S^{(1)}$ is given by \bytaking. Because the metric is flat, we have to
set the first term in \qlt\ equal to zero, and also take into account the
introduction of the cutoff in \actionn\ by modifying the transformations.
The modification needed is similar to \modi:
\eqn\transformers{X^{\mu\si}\!\!\rightarrow \!{X'}^{\mu\si}\!\!=
        X^{\mu\si}\!+b\mintop \tilde{\delta}^{2}_{\Lambda}(\si,\sip)\!
        \left(f_{\mu\nu\lambda}
        (X(\sip))\partial_+ X^{\nu\sip}\partial_- X^{\lambda\sip}\!\!
        + f_{\mu\nu}(X(\sip))\dpm X^{\nu\sip}\!\right).}
 
It is easy to check that, to first order in $b$, \actionn\ is invariant under
\transformers, if at the same time, the fields transform by
\eqn\wehave{\eqalign{ e_{\mu\nu\lambda} &\rightarrow e_{\mu\nu\lambda}
        +2 f_{\mu\nu\lambda}, \cr
        e_{\mu\nu} &\rightarrow e_{\mu\nu}+f_{\mu\nu}+f_{\nu\mu}. \cr}}
Since only the symmetric part of $f_{\mu\nu}$ appears, from now on we will
impose the condition 
\eqn\symf{f_{\mu\nu}=f_{\nu\mu}.}

As we have mentioned earlier, we initially work with flat metric in order
to simplify the exposition. 
After having derived the field equations with the flat metric
as background, we will then show that everything can easily be generalized to
accommodate an arbitrary metric.
 
The above transformations are subject to the condition of unit
determinant (see \deter).
This translates into
\eqn\orwhat{\eqalign{0&=\Tr\log \left( {\delta X' \over \delta X} \right)\cr
        &\cong \mintt \tilde{\delta}^{2}_{\Lambda} (\si,\sip) 
        {\delta \over \delta X^{\mu\si}}
        \left( f_{\mu\nu\lambda}(X(\sip))\partial_+ X^{\nu\sip}
        \partial_- X^{\lambda
        \sip} + f_{\mu\nu}(X(\sip))\dpm X^{\nu\sip}\right).\cr}}
 
Using two dimensional rotational invariance, several cutoff dependent terms
vanish, giving us
\eqn\weget{\tilde{\delta}^{2}_{\Lambda}(0)
\minto \left( \partial_\mu f_{\mu\nu\lambda}\partial_+ X^{\nu\si}
    \partial_- X^{\lambda\si} + \partial_\mu f_{\mu\nu}\dpm X^{\nu\si}
    \right) = 0, }
and
\eqn\andand{f_{\mu\mu}=0,}
where we have used the fact that, by translation invariance, $\tilde{\delta}^
{2}_{\Lambda}(\si,\si)$, which will be shortened to $\diraccutofftilde(0)$, 
does not depend on $\si$.
 
The first condition is satisfied by setting
\eqn\firstgives{\partial_\mu f_{\mu\nu\lambda}-\partial_\lambda\partial_\mu
       f_{\mu\nu}+\partial_\lambda\Lambda_\nu - \partial_\nu\Lambda_\lambda=0,}
where $\Lambda_\mu$ is arbitrary.
It is interesting to identify field combinations that are invariant under
the transformations \wehave, subject to the constraints \andand\ and
\firstgives. $e_{\mu\mu}$ is clearly one such invariant;
another combination which is almost invariant is given by
\eqn\key{k_{\nu\lambda}=\partial_\mu e_{\mu\nu\lambda}-\partial_
        \lambda\partial_\mu
        e_{\mu\nu}.}
Under \wehave, $k_{\nu\lambda}$ undergoes the following gauge transformation:
\eqn\underg{k_{\nu\lambda}\rightarrow k_{\nu\lambda}
    + 2 (\partial_{\nu}\Lambda_{\lambda}-\partial_{\lambda}\Lambda_{\nu}),}
and so it is the appropriate gauge invariant field strength constructed out of
$k_{\nu\lambda}$ that is invariant. Later, we will see that this gauge
invariance is broken for reasons that will become clear.
 
We can now extend the metric given by \metric\ to include the first order
correction in $b$. The key observation is that if there were no restrictions
on the $f$'s, we could gauge away the fields $e_{\mu\nu\lambda}$ and
$e_{\mu\nu}$ by a transformation of the form \transformers\ by setting
\eqn\sett{\eqalign{f_{\mu\nu\lambda}= &\, \ha e_{\mu\nu\lambda}, \cr
    f_{\mu\nu}= &\, \ha e_{\mu\nu}.\cr}}
The metric extended to first order in $b$ is then constructed
starting with flat metric to zeroth order
in $b$ and carrying out the transformation \transformers, with the $f$'s
given by \sett:
\eqn\whereZ{X^{\mu\si}\rightarrow X^{\mu\si}-{b\over2}
        \mintop \tilde{\delta}^{2}_{\Lambda}
        (\si,\sip) \left( e_{\mu\nu\lambda}(X(\sip))
        \partial_+ X^{\nu\sip}\partial_-
        X^{\lambda\sip}+e_{\mu\nu}(X(\sip))\dpm X^{\nu\sip}\right).}
 
The result is
\eqn\itiswhat{\eqalign{g_{\mu\si,\nu\sip} = &\, \eta_{\mu\nu}
        \diraccutoff(\si,\sip) + b h^{(1)}_{\mu\si,\nu\sip},
         \cr  h^{(1)}_{\mu\si,\nu\sip}= & -{1\over2}{\delta\over \delta
        X^{\nu\sip}} \left(
        e_{\mu\alpha\lambda}(X(\si))\partial_+ X^{\alpha\si} \partial_-
        X^{\lambda\si} + e_{\mu\alpha}(X(\si))\partial_+\partial_-
        X^{\alpha\si} \right) \cr
        & +(\mu\si \leftrightarrow \nu\sip). \cr}}
 
If the constraints \andand\ and \firstgives\ did not exist, this would be a
trivial metric, equivalent to a flat metric. In that case, there would be no
need to go to the trouble of constructing it; it 
would have been simpler to fix
gauge by eliminating the fields $e_{\mu\nu\lambda}$ and $e_{\mu\nu}$.
However,  the constraints on the $f$'s make \itiswhat\ a non-trivial metric:
Because of these constraints, $h_{\mu\si,\nu\sip}$ can no longer be
transformed away, and neither can the $e$'s be completely eliminated. It is
easy to check directly, using \wehave, that, even in the presence of the
constraints, \itiswhat\ transforms correctly to first order in $b$ under
\transformers.
 
From the metric given above, one can find the first order correction in $b$ to
the connection and the generators of the conformal transformations (see
\specc\ and the related discussion). The standard formula of differential
geometry expressing the connection in terms of metric gives
\eqn\andgamma{\eqalign{\Gamma^{(1)\lambda\sippp}_{\mu\si,\nu\sip}\!= &
        -{1\over2}  {\delta^2
        \over \delta X^{\mu\si}\delta X^{\nu\sip}}\mintopp \tilde{\delta}
        ^{2}_{\Lambda}(\sippp,\sipp)\cr &\qquad\times\left( 
        e_{\lambda\alpha\beta}(X(\sipp))\partial_+
        X^{\alpha\sipp}\partial_-X^{\beta\sipp}\!+e_{\lambda\alpha}(X(\sipp))
        \partial_+\partial_- X^{\alpha\sipp}\right).\cr} }
 
Now let us compute the corrected conformal generators. Write \specc\ as
$$
\eqalign{ F^{\mu\si}_{v}(X)= & \mintop v(\sipa) F^{\mu\si}_{\sip}(X),\cr
	F^{\mu\si}_{\sip}(X)= &\, \dirac (\sip -\si)\partial_{+}X^{\mu\si}+
 	f^{\mu\si}_{\sip}(X),\cr}
$$
where $f^{\mu\si}_{\sip}$ starts at first order in $b$. Taking advantage of the
fact that $F$ transforms like a vector in the indices $\mu\si$, the first order
correction is computed exactly as in the case of the metric. Start with
$$
F^{\mu\si}_\sip (X)=\dirac (\sip-\si)\partial_+ X^{\mu\si}
$$
at zeroth order, and apply the vector transformation law to it under
coordinate transformation \whereZ, with the result
\eqn\Fmore{\eqalign{\mintop v(\sipa)f^{\mu\si}_\sip
        =&-{b\over2}\mintop \partial_\sia \tilde{\delta}^{2}_{\Lambda}
(\si,\sip)(v(\sia)- v(\sipa))\cr
        &\qquad\times\left(e_{\mu\nu\lambda}(X(\sip))\partial_+X^{\nu\sip}
        \partial_-X^{\lambda
        \sip}+e_{\mu\nu}(X(\sip))\dpm X^{\nu\sip}\right). \cr}}
This result can be simplified in the limit of large cutoff. As $\Lambda$
becomes large, $\diraccutoff (\si,\sip) \rightarrow \dirac (\si-\sip)$, and
$\partial_\sia \tilde{\delta}^{2}_{\Lambda} (\si,\sip)
 (v(\sia)-v(\sipa)) \rightarrow
-\tilde{\delta}^{2}_{\Lambda} (\si,\sip) v'(\sipa)$, so we can write
\eqn\writemore{\eqalign{\mintop v(\sipa) f^{\mu\si}_\sip \cong & {b\over2}
        \mintop v'(\sipa)
        \tilde{\delta}^{2}_{\Lambda} (\si,\sip)\cr
        &\qquad\times\left(e_{\mu\nu\lambda}(X(\sip))
        \partial_+X^{\nu\sip}
        \partial_-X^{\lambda\sip}+e_{\mu\nu}(X(\sip))
        \dpm X^{\nu\sip}\right).\cr}}
 
We have now at hand all the information  needed to evaluate \basic\ to first
order in $b$; the action is given by \actionn, the connection by \andgamma,
and the conformal generator by \Fmore. Since the calculation is straightforward
but somewhat tedious, we skip the details and instead, indicate the main steps.
Part of the calculation was already carried out for the non-covariant case in
section 4; all we have to do is to add the extra terms that arise from the
connection and from $f$ in \Fmore. We first calculate the terms that contribute
to $E_{G}$ in  \basic; a simple calculation gives
\eqn\another{\mintop v(\sipa)f^{\mu\si}_\sip {\delta S^{(0)}
        \over \delta X^{\mu\si}}
        =-b \minto v'(\sia)\dpm X^{\mu\si}\left(e_{\mu\nu\lambda}\partial_+
        X^{\nu\si}\partial_-X^{\lambda\si}+e_{\mu\nu}\dpm X^{\nu\si}
        \right),}
and therefore, to first order in $b$,
\eqn\neednames{\left(\mintop v(\sipa) F^{\mu\si}_\sip {\delta S \over
        \delta X^{\mu\si}}\right)^{(1)} \cong
        \minto v'(\sia)\left(e_{\m1\m2,\n1\n2}\dpmu1\dpmu2\dmnu1\dmnu2
        \right),}
since the other two terms, $e_{\m1\m2\m3}$ and $e_{\m1\m2}$, cancel. Next,
expanding the Tr log as in \tlG, we  compute $H_{\mu\si,\nu\sip}$ to
first order in $b$:
\eqn\compute{\eqalign{H_{\mu\si,\nu\sip}^{(1)}= &
        {\dirac S^{(1)} \over \delta X^{\mu\si}
        \delta X^{\nu\sip}}-
        \Gamma^{\lambda\sipp}_{\mu\si,\nu\sip}{\delta S^{(0)}\over \delta
        X^{\lambda\sipp}}\cr
        = &{\dirac S' \over \delta X^{\mu\si} \delta X^{\nu\sip}}+
         \partial_\sia\partial_\sis
        \left({\delta \over \delta X^{\nu\sip}}
        \left(e_{\mu\alpha\beta}\partial_+X^{\alpha\si}\partial_-X^{\beta\si}+
        e_{\mu\alpha}\dpm X^{\alpha\si}\right)\right)\cr
        &\null+(\mu\si \leftrightarrow  \nu\sip), \cr}}
where,
\eqn\Sprime{S'=\minto \left( e_{\m1\m2,\n1\n2}\dpmu1\dpmu2\dmnu1\dmnu2
        \right).}
Next, we apply $\delta$ (see \var) to $H_{\mu\si,\nu\sip}$.
The contribution coming from the first term on the right in \compute,
$$
\left(\mintopp v(\sipp)\partial_+ X^{\lambda\sipp}{\delta
        \over \delta X^{\lambda\sipp}}+\delta_\Lambda \right)
        \left(\eta^{\mu\nu}\Delta^{\si,\sip}(\Lambda)
        {\dirac S' \over \delta X^{\mu\si} \delta X^{\nu\sip}} \right),
$$
has  already been calculated in the last section; it is given by setting
$e_{\m1\m2\m3}$ and $e_{\m1\m2}$ in  \soso\ equal to zero. The contribution
of the second term in \compute, after a somewhat lengthy computation, is
given by
\eqn\addingup{\eqalign{&\ha b \left(\mintopp v(\sippa)\partial_+ 
        X^{\lambda\sipp}
        {\delta \over
        \delta X^{\lambda\sipp}}+\delta_\Lambda\right)
        \eta^{\mu\nu}\Delta^{\si,\sip}(\Lambda)\cr
        &\qquad\times\left(\partial_\sia\partial_\sis
        {\delta\over\delta X^{\nu\sip}}
        \left(e_{\mu\alpha\beta}
        \partial_+X^{\alpha\si}\partial_-X^{\beta\si}+e_{\mu\alpha}
        \dpm X^{\alpha\si}
        \right)+(\mu\si\leftrightarrow\nu\sip)\right)=\cr
        &\quad=-b\tilde{\delta}^{2}_{\Lambda}(0)
        \minto v'(\sia)\partial_+X^\alpha
        \partial_-X^\beta
        \left(\partial_\mu e_{\mu\alpha\beta}-\partial_\beta\partial_\mu
        e_{\mu\alpha}
        \right)- b\dpm\tilde{\delta}^{2}_{\Lambda}(0)
        \minto v'(\sia)e_{\mu\mu},\cr}}
with 
$$
\dpm\tilde{\delta}^{2}_{\Lambda}(0)\equiv \left(\partial_\sia
\partial_\sis\tilde{\delta}^{2}_{\Lambda}(\si,\sip)\right)_{\si =\sip}.
$$

The main steps in the computation are the following: The critical term to be
evaluated turns out to be
\eqn\firststep{\eqalign{&\mintt (\delta_{\Lambda}\Delta^{\si,\sip}(\Lambda))
        \left(\partial_\sia\partial_\sis
        {\delta\over\delta X^{\mu\sip}}
        \left(e_{\mu\alpha\beta}
        \partial_+X^{\alpha\si}\partial_-X^{\beta\si}+e_{\mu\alpha}
        \dpm X^{\alpha\si} \right)\right)= \cr
        & =\! - \!\minto\!\! \left(\!(\delta_{\Lambda}
        \tilde{\delta}^{2}_{\Lambda}
        (\si,\sip))_{\si=\sip}\partial_{\mu}e_{\mu\alpha\beta} 
        \partial_{+}X^{\alpha\si}\partial_{-}X^{\beta\si}\!\! + 
        (\partial_\sia\partial_\sis
        \delta_{\Lambda}\tilde{\delta}^{2}_{\Lambda}(\si,\sip))_{\si=\sip}
        e_{\mu\mu}\dpm X^{\alpha\si}\! \right)\!,\cr}}
and the cutoff dependent factors can be simplified using \invflat:
\eqn\secondstep{\left(\delta_{\Lambda}\tilde{\delta}^{2}_{\Lambda}(\si,\sip)
        \right)_{\si=\sip}= v'(\sia) \tilde{\delta}^{2}_{\Lambda}(0).}

To obtain $E_{G}$ to first order,
the above correction term should be added to \soso, with $e_{\m1\m2\m3}$ and
$e_{\m1\m2}$ set equal to zero.
 
We now consider the term $E_{\cal M}$ in \basic; a straightforward calculation
gives the result
\eqn\detterm{\eqalign{E^{(1)}_{\cal M}& =
        \ha\mintopp v(\sippa)\!\left(\!{\delta F^{\mu\si}_\sipp\over  \delta
        X^{\alpha\sippp}}G^{\alpha\sippp,\nu\sip}+{\delta F^{\nu\sip}_\sipp\over
        \delta X^{\alpha
        \sippp}}G^{\mu\si,\alpha\sippp}-\delta_\Lambda
        G^{\mu\si,\nu\sip}\!\right)\!\!\left(
        {\dirac\Tr\log {\cal M}\over\delta P^{\mu\si}\delta P^{\nu\sip}}
        \right)_{P=0}\cr
        &=\ha\left(\left(v(\sia)\partial_\sia+v(\sipa)
        \partial_\sipa\right)\Delta^
        {\mu\si,\nu\sip}-\delta_\Lambda\Delta^{\mu\si,\nu\sip}\right)\left(
        {\dirac\Tr\log {\cal M}\over\delta P^{\mu\si}\delta P^{\nu\sip}}
        \right)_{P=0}\cr
        &={1\over 24\pi}\mintt{v(\sia)-v(\sipa)\over \sia-\sipa}\eta^{\mu\nu}
        \left({\delta\Gamma^{\lambda\sipp}_
        {\lambda\sipp,\mu\si}\over
        \delta X^{\nu\sip}}+{\delta\Gamma^{\lambda\sipp}_
        {\lambda\sipp,\nu\sip}\over
        \delta X^{\mu\si}}+{\delta\Gamma^{\lambda\sipp}_{\mu\si,\nu\sip}\over
        \delta X^{\lambda\sipp}}\right)\cr
        &=-{b\over16\pi}\minto v'(\sia)\Big(\dpm\diraccutofftilde(0)
        \bbox e_{\lambda
        \lambda}\cr &\qquad\qquad\null+\diraccutofftilde(0)
        \left(\bbox k_{\nu\lambda}-
        \partial_\nu\partial_\mu
        k_{\mu\lambda}+\partial_\lambda\partial_\mu k_{\mu\nu}\right)
        \partial_+X^\nu
        \partial_-X^\lambda\Big),\cr}}
where $\Gamma$ is given by \andgamma, and $k$ is defined by \key.
 
Putting together \soso, \addingup\ and \detterm\ in \basic, we finally get the
equations for the first massive level. These equations contain cutoff
independent terms, which come only from \soso, and cutoff dependent terms,
which all come from \addingup\ and \detterm. We note that all the cutoff
dependent contributions come from terms proportional to the connection
$\Gamma$, and therefore they are absent from a non-covariant calculation.
We first write down the cutoff independent equations:
\eqn\lastone{\eqalign{\bbox e_{\m1\m2,\n1\n2}+  16 \pi e_{\m1\m2,\n1\n2}&
        =0, \cr
        e_{\mu\mu,\n1\n2}=0,\quad\qquad e_{\m1\m2,\nu\nu}& =0,\cr
        \partial_\mu e_{\mu\m1,\n1\n2}=0,\quad \partial_\nu
        e_{\m1\m2,\nu\n1}& =0.\cr}}
 
We have one equation of motion and four constraints. In addition, we have 
three cutoff dependent equations. Two of them follow
from the conformal transformations in $\sia$:
\eqn\cutoffdep{\eqalign{{1\over16\pi}\left(\bbox k_{\nu\lambda}-
        \partial_\nu\partial_
        \mu k_{\mu\lambda}+\partial_\lambda\partial_\mu
        k_{\mu\nu}\right)+k_{\nu
        \lambda}&=0, \cr
        {1 \over 16 \pi} \bbox e_{\lambda\lambda}+ e_{\lambda\lambda}&= 0,\cr}}
where $k$ is defined by \key. The remaining
equation (there is a fourth, repeated equation, for $e_{\lambda\lambda}$),
results from conformal transformations in $\sis$ and it is
conveniently written in terms of a field $\bar{k}$, defined by
\eqn\keybar{\bar{k}_{\nu\lambda}=\partial_{\mu} e_{\mu\nu\lambda} -
        \partial_{\nu}\partial_{\mu} e_{\mu\lambda},}
and it reads
\eqn\cutoffdepp{{1 \over 16\pi}\left(
        \bbox \bar{k}_{\nu\lambda} - \partial_{\lambda}
        \partial_{\mu}\bar{k}_{\nu\mu}+ \partial_{\nu}\partial_{\mu}
        \bar{k}_{\lambda\mu}\right) + \bar{k}_{\nu\lambda}= 0.}
 
The equation satisfied by $k$ is not invariant under the gauge transformations
given by \underg. The reason for this is the following: In the computation of 
the determinant, the cutoff dependent factor $\tilde{\delta}^{2}_{\Lambda}
(\si,\sip)$  at $\si=\sip$
is $\si$ independent and therefore
it can be put in front of the integral in \weget. The integral itself is then
invariant under the gauge transformation \underg. On the other hand, in the
main step leading to \addingup, the cutoff variation of the same factor
at $\si=\sip$ is $\si$ dependent (see \firststep, \secondstep, and also
the discussion following \invflat). As a consequence, an additional factor
$v'(\sia)$, as compared to \weget, appears in the integral on the right
hand side of \addingup, and this spoils gauge invariance under \underg. It is,
therefore, necessary to modify the condition \firstgives; it should be replaced 
by
\eqn\replaced{\eqalign{\partial_{\mu}f_{\mu\nu\lambda}&=0,\cr
\partial_{\mu}f_{\mu\nu}&=0.\cr}}

Both $k$ and $\bar{k}$ are invariant under the transformations satisfying these
more stringent conditions.

Going back to the equations \lastone, we see that two of the constraints are
too stringent,
\eqn\toostr{e_{\mu\mu,\n1\n2}=0, \qquad e_{\m1\m2,\nu\nu}=0,}
eliminating  degrees of freedom from the field $e_{\m1\m2,\n1\n2}$ which are
present in the string spectrum (see Appendix B). The hope is that $k$ and 
$\bar{k}$ could supply the missing degrees of freedom. We shall see below that
this happens in the left-right symmetric case,  with parity invariance on
the world sheet, which interchanges $\sia$ and $\sis$. In this case, $e$ is
invariant under the interchange of the $\mu$'s with $\nu$'s, and the components
eliminated by \toostr\ are the same as those of a symmetric second rank tensor.
We have analyzed equations \cutoffdep\ and \cutoffdepp\ in the left-right
symmetric case, when
$$
e_{\mu\nu\lambda}=e_{\mu\lambda\nu}.
$$
Defining
$$
l_{\nu\lambda}\equiv\partial_{\mu}e_{\mu\nu\lambda}, \qquad
l_{\nu}\equiv\partial_{\mu} e_{\mu\nu},
$$
and
$$
A_{\mu\nu}\equiv 2 l_{\mu\nu}- \partial_{\mu}l_{\nu}- \partial_{\nu}l_{\mu},
\qquad L_{\mu\nu}\equiv\partial_{\mu}l_{\nu}- \partial_{\nu} l_{\mu},
$$
one can easily show that
equations \cutoffdep\ and \cutoffdepp\ are equivalent to the equations
\eqn\equival{\eqalign{{1 \over 16 \pi}\bbox A_{\mu\nu} + A_{\mu\nu}&=0, \cr
{1 \over 16 \pi} \bbox L_{\mu\nu} + L_{\mu\nu}&=0, \cr}} 
plus the constraint
\eqn\restr{\partial_{\mu} \partial_{\nu}A_{\mu\lambda} -\partial_{\mu}
\partial_{\lambda}A_{\mu\nu}= \bbox L_{\lambda\nu}.}

The number of independent degrees of freedom of the above system is the
same as that of a symmetric second order tensor minus a scalar. The missing
scalar is provided by $e_{\mu\mu}$, so in the final count, the fields $k$ and
$\bar{k}$ provide the missing degrees of freedom needed to establish agreement
with the string theory spectrum. Unfortunately, in the general case with no
left-right symmetry, there are still missing degrees of freedom, and at the
present time, we have no solution to this problem. Our suspicion is that our
method in its present form is applicable only in the symmetric case, and some
new ideas are needed to extend it to the general case.

We close this section by a brief description of the promised extension of the
results of this section to the case of a general gravitational background. This
means replacing the flat background given by
$\eta_{\mu\nu} \diraccutoff(\si,\sip)$ in \itiswhat\ by the metric
$g_{\mu\si,\nu\sip}$ of \metric. We have to show that the equations of this
section can be covariantized with respect to this metric. Most of the time, the
task is trivial; one has to keep the upper and lower indices of tensors
match correctly and use the metric to raise and lower indices as needed. For
example, in  \bytaking, the first term on the right is correctly
written, since $\partial_+ X^\mu$ and $\partial_- X^\mu$ transform 
as contravariant vectors. On the
other hand, $\dpm X^\mu$ is not a vector; it should be replaced by
$$
\dpm X^{\mu} \rightarrow \dpm X^{\mu}+ \partial_+ X^\alpha 
\partial_- X^\beta \Gamma^{\mu}_{\alpha\beta}(X(\si)),
$$
where the connection $\Gamma$ is given by \connect. Similarly, the partial
derivative with respect to $X$ in \itiswhat\ should be replaced by the covariant
derivative using the same connection: For example, 
$$
{\delta \over \delta X^{\mu\si}}(V_{\nu\sip}) \rightarrow 
{\delta \over \delta X^{\mu\si}}(V_{\mu\sip}) - 
\Gamma^{\lambda\sipp}_{\mu\si,\nu\sip} V_{\lambda\sipp},
$$
for a vector $V_{\nu\sip}$. One can easily show that everything in this section
goes through with  these modifications. Notice, however, that in all this we
have worked only with the metric, which is symmetric, and we have dropped the
antisymmetric tensor altogether. This is clearly permissible only in a
left-right symmetric model. It is clear that, in order to generalize our
treatment to the left-right non-symmetric string, we have to figure out how
to incorporate the antisymmetric tensor in the discussion above.

\newsec{Conclusions}

In this paper we have proposed a new approach for deriving the string field
equations from a general sigma model on the world sheet. Those equations
can be made covariant under not only local, but also non-local transformations
in the field space. In this approach the world sheet
one loop result is exact, although
it may only give incomplete information, to be supplemented by higher loop
results.
We applied this method to derive the equations for the tachyon, massless and
first massive level. The spectrum of states that follows from the linear
part of these equations of motion was shown to agree with the known spectrum
of strings. This is in contrast with a non-covariant approach, where the
equations are too weak to produce the  right spectrum.

In this paper we only analyzed the linear part of the equations. We did not
address the question of string interactions, neither did we attempt to extend
our results to higher string loops.
It would be
desirable to go beyond the expansion we have used, and establish exact
covariance under non-local transformations. Also,  natural generalizations
such as a better treatment of left-right non-symmetric closed string, strings
with boundaries (open strings) 
and fermionic strings are worthy of investigation.

\bigbreak\bigskip\bigskip\centerline{{\bf Acknowledgements}}\nobreak

We are grateful to John Ellis and Nathan Berkowitz for help with references.


\appendix{A}{}

In this section we fill up the steps that lead from \general\ to \bytaking.
As it was said in the comments that follow \general, of the eight fields
present there, all but three can be eliminated by linear gauge transformations.
The six distinct linear gauge transformations that we can add to \general\ are:
\eqn\canadd{\eqalign{1)\quad\partial_+(&\vep_{\mu,\n1\n2}\partial_+X^\mu\dmnu1
        \dmnu2) = \partial_{\m1} \vep_{\m2,\n1\n2}\dpmu1\dpmu2\dmnu1\dmnu2 \cr
        &+ \vep_{\mu,\n1\n2}(\partial^2_+X^\mu\dmnu1\dmnu2 +2\partial_+
        X^\mu\dpm X^{\n1}\dmnu2) \cr
        2)\quad\partial_-(&{\bar\vep}_{\m1\m2,\nu}\dpmu1\dpmu2\partial_-
        X^\nu)=\partial_{\n1}{\bar\vep}_{\m1\m2,\n2}\dpmu1\dpmu2\dmnu1\dmnu2
        \cr &+{\bar\vep}_{\m1\m2,\nu}(2\dpmmu1\dpmu2\partial_- X^\nu +
        \dpmu1\dpmu2\partial^2_- X^\nu) \cr
        3)\quad\partial_+(&\vep^{(1)}_{\m1\m2}\dpmmu1\dmmu2)=\partial_\nu
        \vep^{(1)}_{\m1\m2}\partial_+ X^\nu \dpmmu1\dmmu2 \cr & + \vep^{(1)}_
        {\m1\m2}(\dppmmu1\dmmu2+\dpmmu1\dpmmu2) \cr
        4)\quad\partial_-(&{\bar\vep}^{(1)}_{\m1\m2}\dpmmu1\dpmu2)=
        \partial_\nu {\bar\vep}^{(1)}_{\m1\m2}\partial_- X^\nu \dpmmu1\dpmu2
        \cr &+{\bar\vep}^{(1)}_{\m1\m2}(\dmmpmu1\dpmu2+\dpmmu1\dpmmu2) \cr
        5)\quad\partial_+(&\vep^{(2)}_{\m1\m2}\dpmu1\dmmmu2)=\partial_\nu
        \vep^{(2)}_{\m1\m2}\partial_+ X^\nu \dpmu1\dmmmu2 \cr
        &+\vep^{(2)}_{\m1\m2}(\dppmu1\dmmmu2+\dpmu1\dpmmmu2) \cr
        6)\quad\partial_-(&{\bar\vep}^{(2)}_{\m1\m2}\dmmu1\dppmu2)=
        \partial_\nu {\bar\vep}^{(2)}_{\m1\m2}\partial_- X^\nu \dmmu1\dppmu2
        \cr &+{\bar\vep}^{(2)}_{\m1\m2}(\dmmmu1\dppmu2+\dmmu1\dppmmu2). \cr}}
To eliminate $e^{(4)}_{\mu\nu}$ choose
\eqn\elimi{e^{(4)}_{\m1\m2}+\vep^{(2)}_{\m1\m2}+{\bar\vep}^{(2)}_{\m2\m1}=0.}
To eliminate $e^{(2)}_{\mu\nu}$ choose
\eqn\elimii{e^{(2)}_{\m1\m2}+\vep^{(1)}_{\m1\m2}+{\bar\vep}^{(2)}_{\m2\m1}=0.}
To eliminate $e^{(3)}_{\mu\nu}$ choose
\eqn\elimiii{e^{(3)}_{\m1\m2}+{\bar\vep}^{(1)}_{\m1\m2}+\vep^{(2)}_{\m2\m1}=0.}
To eliminate $e^{(2)}_{\mu\nu\lambda}$ choose
\eqn\elimiv{e^{(2)}_{\m1\m2\m3}+\vep_{\m1,\m2\m3}+\ha(\partial_\m2
        {\bar\vep}^{(2)}_{\m3\m1}+\partial_\m3{\bar\vep}^{(2)}_{\m2\m1})=0.}
To eliminate $e^{(3)}_{\mu\nu\lambda}$ choose
\eqn\elimv{e^{(3)}_{\m1\m2\m3}+{\bar\vep}_{\m2\m3,\m1}+\ha(\partial_\m2
        \vep^{(2)}_{\m3\m1}+\partial_\m3\vep^{(2)}_{\m2\m1})=0.}

By choosing $\vep^{(2)}_{\m1\m2}+{\bar\vep}^{(2)}_{\m2\m1}$, $\vep^{(1)}
        _{\m1\m2}$, ${\bar\vep}^{(1)}_{\m1\m2}$, $\vep_{\m1,\m2\m3}$, and
        ${\bar\vep}_{\m2\m3,\m1}$ properly, we can eliminate everything
except $e_{\m1\m2,\n1\n2}$, $e_{\m1\m2\m3}$ and $e_{\m1\m2}$; we dropped
the superscript  (1) after gauge fixing.
The transformations that preserve this gauge are
\eqn\preserve{\eqalign{
              \vep^{(2)}_{\m1\m2}+{\bar\vep}^{(2)}_{\m2\m1}=0, \qquad
              \vep^{(1)}_{\m1\m2}&+{\bar\vep}^{(2)}_{\m2\m1}=0, \qquad
              {\bar\vep}^{(1)}_{\m1\m2}+\vep^{(2)}_{\m2\m1}=0, \cr
              2\vep_{\m1,\m2\m3}+\partial_\m2 & {\bar\vep}^{(2)}_{\m3\m1}
              +\partial_\m3 {\bar\vep}^{(2)}_{\m2\m1}=0, \cr
              2{\bar\vep}_{\m2\m3,\m1}+\partial_\m2& \vep^{(2)}_{\m3\m1}
              +\partial_\m3 \vep^{(2)}_{\m2\m1}=0, \cr }}
plus, we could also add
\eqn\alsoadd{\partial_+(\vep_\m1\dpmmmu1)-\partial_-(\vep_\m1\dmppmu1)=
        \partial_\m1\vep_\m2(\dpmu1\dpmmmu2-\dmmu1\dmppmu2).}
All of these linear transformations act trivially on \bytaking; they
 leave $e_{\m1\m2,\n1\n2}$, $e_{\m1\m2\m3}$
and $e_{\m1\m2}$ invariant. This means that linear gauges
are completely fixed in the form given by  \bytaking.


\appendix{B}{}

In this section we show that the constraints (equations \cconst\ and \fromvp)
obtained in section 4 for the first massive level fields are too weak. This
will be done by comparing those constraints with the analogue constraints
of the first massive level of the string.

To obtain these consider the most general level 2 state $|s\rangle$ given by
\eqn\state{\eqalign{|s\rangle =& \left( E_{\m1\m2,\n1\n2}a^{\dag\m1}_1
        a^{\dag\m2}_1 b^{\dag\n1}_1 b^{\dag\n2}_1 +E_{\mu,\n1\n2}
        a^{\dag\mu}_2 b^{\dag\n1}_1 b^{\dag\n2}_1 \right. \cr
        & \left. \null+E_{\m1\m2,\nu} a^{\dag\m1}_1 a^{\dag\m2}_1 b^{\dag\nu}_2
        + E_{\mu\nu} a^{\dag\mu}_2 b^{\dag\nu}_2 \right) |0\rangle , \cr }}
where $a^\mu_n$, $a^{\dag\mu}_n$ and $b^\mu_n$, $b^{\dag\mu}_n$ are the
closed string operators. 
This state satisfies the relations
$$
(L_0-1)|s\rangle=({\bar L}_0-1)|s\rangle=0 
$$
and
$$
L_1|s\rangle ={\bar L}_1|s\rangle=0,
\qquad\quad
L_2|s\rangle ={\bar L}_2|s\rangle=0,
$$
from which we get some conditions between the $E$'s. The gauge freedom 
present in these conditions can be taken care of by adding zero norm
states to $|s\rangle$. After that is done we get the constraints
\eqn\previous{\eqalign{
        p^\mu E_{\mu\m1,\n1\n2}+\sqrt{2}p_\m1 E_{,\n1\n2}&=0, \cr
        p^\nu E_{\m1\m2,\n1\nu}+\sqrt{2}p_\n1 E_{\m1\m2,}&=0, \cr
        p^\mu E_{\mu\m1,}+\sqrt{2}p_\m1 E &=0, \cr
        p^\nu E_{,\n1\nu}+\sqrt{2}p_\n1 E &=0, \cr
        4\sqrt{2} E_{,\n1\n2}-E_{\mu\mu,\n1\n2} &=0, \cr
        4\sqrt{2} E_{\m1\m2,}-E_{\m1\m2,\nu\nu} &=0, \cr}}
where the new $E$'s are related to the old ones by
$$
E_{\mu,\n1\n2}=p_\mu E_{,\n1\n2},\qquad E_{\m1\m2,\nu}=
        p_\nu E_{\m1\m2,},\qquad E_{\mu,\nu}=p_\mu p_\nu E.
$$
Combining, the only constraints on $E_{\m1\m2,\n1\n2}$ are
\eqn\arewhat{\eqalign{
        p^\mu E_{\mu\m1,\n1\n2}+{1\over4}p_\m1 E_{\mu\mu,\n1\n2} &=0, \cr
        p^\nu E_{\m1\m2,\n1\nu}+{1\over4}p_\n1 E_{\m1\m2,\nu\nu} &=0, \cr }}
and all other $E$'s are given in terms of $E_{\m1\m2,\n1\n2}$. 
If we consider that the $E$'s play the analogue roll of the $e$'s in section
4, then the constraints \cconst\ and \fromvp\ are not powerfull enough,
because for example, $e_{\m1\m2\m3}$ and $e_{\m1\m2}$ are not completely
determined in terms of $e_{\m1\m2,\n1\n2}$.

\listrefs
\bye